\documentclass[twocolumn]{aastex631}
\usepackage{rotating}

\received{31 January 2023}
\revised{20 March 2023}
\accepted{23 March 2023}
\submitjournal{AAS Journals}

\shorttitle{IRMP Stars and the Astrophysics of Thermonuclear Events. I.}
\shortauthors{Reggiani et al.}

\begin{document}

\title{Iron-rich Metal-poor Stars and the Astrophysics of Thermonuclear
Events Observationally Classified as Type Ia Supernovae. I. Establishing
the Connection}

\correspondingauthor{Henrique Reggiani}
\email{hreggiani@carnegiescience.edu}

\author[0000-0001-6533-6179]{Henrique Reggiani}
\altaffiliation{Carnegie Fellow}
\affiliation{The Observatories of the Carnegie Institution
for Science, 813 Santa Barbara St, Pasadena, CA 91101, USA}

\author[0000-0001-5761-6779]{Kevin C.\ Schlaufman}
\affiliation{William H.\ Miller III Department of Physics \& Astronomy,
Johns Hopkins University, 3400 N Charles St, Baltimore, MD 21218, USA}

\author[0000-0003-0174-0564]{Andrew R.\ Casey}
\affiliation{School of Physics \& Astronomy, Monash University, Wellington
Road, Clayton 3800, Victoria, Australia}
\affiliation{ARC Centre of Excellence for All Sky Astrophysics in 3
Dimensions(ASTRO 3D), Canberra, ACT 2611, Australia}

\begin{abstract}

\noindent
The progenitor systems and explosion mechanisms responsible for the
thermonuclear events observationally classified as Type Ia supernovae
are uncertain and difficult to uniquely constrain using traditional
observations of Type Ia supernova host galaxies, progenitors, light
curves, and remnants.  For the subset of thermonuclear events that are
prolific producers of iron, we use published theoretical nucleosynthetic
yields to identify a set of elemental abundance ratios infrequently
observed in metal-poor stars but shared across a range of progenitor
systems and explosion mechanisms: $[\text{Na,Mg,Co/Fe}]<0$.  We label
stars with this abundance signature ``iron-rich metal-poor'' or IRMP
stars.  We suggest that IRMP stars formed in environments dominated by
thermonuclear nucleosynthesis and consequently that their elemental
abundances can be used to constrain both the progenitor systems
and explosion mechanisms responsible for thermonuclear explosions.
We identify three IRMP in the literature and homogeneously infer their
elemental abundances.  We find that the elemental abundances of BD+80
245, HE 0533--5340, and SMSS J034249.53--284216.0 are best explained
by the (double) detonations of sub-Chandrasekhar mass CO white dwarfs.
If our interpretation of IRMP stars is accurate, then they should be very
rare in globular clusters and more common in the Magellanic Clouds and
dwarf spheroidal galaxies than in the Milky Way's halo.  We propose that
future studies of IRMP stars will quantify the relative occurrences of
different thermonuclear event progenitor systems and explosion mechanisms.

\end{abstract}


\section{Introduction}\label{intro}

Several lines of observational evidence support the conclusion that
Type Ia supernovae are produced by the thermonuclear explosions of
carbon-oxygen (CO) white dwarfs left behind as the embers of $1~M_{\odot}
\lesssim M_{\ast} \lesssim 8~M_{\odot}$ stars \citep[e.g.,][]{whelan1973}.
Unlike Type Ib/c and Type II supernovae that are only observed in
star-forming galaxies and thought to result from the explosions of
massive stars, Type Ia supernovae are observed in both star-forming
and passively evolving galaxies \citep{zwicky1958}.  The lack of
hydrogen and helium lines in the spectra of Type Ia supernovae point
to degenerate objects, while lines of silicon, calcium, and iron
blueshifted by $10^{4}$ km s$^{-1}$ indicate explosions and the presence
of significant amounts of both intermediate-mass and iron-peak elements
\citep[e.g.,][]{hoyle1960,branch1983,branch1985,kirshner1993,mazzali1993}.
The fusion of carbon and oxygen at high absolute densities
and temperatures can produce at relatively low densities
intermediate-mass elements and at relatively high densities
iron-peak elements like radioactive $^{56}$Ni in nuclear
statistical equilibrium.  The energy thereby released is
sufficient to unbind a CO white dwarf with $M_{\text{WD}} \approx
1~M_{\odot}$ and accelerate a typical Type Ia supernova ejecta
mass to $10^{4}$ km s$^{-1}$ \citep[e.g.,][]{arnett1969,maoz2014}.
The characteristic few week rise to maximum light followed by
an order of magnitude decline in one month and then less rapid
decline thereafter is well explained by the radioactive decays
$^{56}\text{Ni}\rightarrow$$^{56}\text{Co}\rightarrow$$^{56}\text{Fe}$
\citep[e.g.,][]{pankey1962,colgate1969}.  Observations of Galactic
supernovae remnants left behind by Type Ia supernovae support this
picture \citep[e.g.,][]{minkowski1964,badenes2006}.

There are now highly refined models of several progenitor systems and
explosions mechanisms that are broadly consistent with observed Type
Ia supernovae host galaxies, progenitors, light curves, and remnants
\citep[e.g.,][]{hillebrandt2013}.  While this broad model consistency with
observational data represents an impressive achievement of theoretical
astrophysics, no single progenitor system and explosion mechanism
explains all Type Ia supernovae \citep[e.g.,][]{hoeflich1996,nugent1997}.
Most models invoke thermonuclear explosions of CO white dwarfs due
to interactions with binary companions.  In the single-degenerate
scenario, a thermonuclear supernova results from a CO white dwarf's
binary interactions with a star leaving the main sequence, on the giant
branch, or that has lost its hydrogen envelope.  In the double degenerate
scenario, a thermonuclear supernova results from a CO white dwarf's
interactions with another white dwarf.  In either the single-degenerate or
double-degenerate scenarios, the CO white dwarf producing a thermonuclear
explosion may be near the Chandrasekhar mass $1.3~M_{\odot} \lesssim
M_{\text{Ch}} \lesssim 1.4~M_{\odot}$ \citep[e.g.,][]{leung2018} or
significantly below the Chandrasekhar mass (i.e., sub $M_{\text{Ch}}$).

Preceded by approximately 100 years of
convective carbon burning usually called simmering
\citep[e.g.,][]{nomoto1984,piro2008a,piro2008b,piro2008c,schwab2017},
the thermonuclear explosions of near-$M_{\text{Ch}}$ CO white dwarfs
accreting hydrogen can in principle proceed as pure detonations,
pure deflagrations, or delayed detonations.  Pure detonations are
thought to produce copious amounts of iron-peak elements but too
little intermediate-mass elements to explain the lines of those
species in Type Ia supernovae \citep{arnett1971}.  They are also too
bright to explain most Type Ia supernovae light curves.  In a pure
deflagration, carbon fusion ignited in the interior of a white dwarf
moves toward the surface as the energy released is conducted outward by
the degenerate electron gas \citep{nomoto1976,nomoto1984}.  The energy
causes the white dwarf to expand, allowing for the production of both
iron-peak elements in dense regions and intermediate-mass elements in
less dense regions.  In a delayed detonation, subsonic transport of
energy outward from the deflagration causes an expansion of the white
dwarf before a following supersonic detonation wave unbinds the white
dwarf and produces nucleosynthesis that does not reach the iron peak
in relatively low-density environments \citep{khokhlov1991}.  This
delayed detonation can occur as a deflagration-to-detonation transition
\citep[DDT;][]{gamezo2003,gamezo2004,gamezo2005,ropke2007,bravo2008,seitenzahl2011},
a gravitationally confined detonation
\citep[GCD;][]{plewa2004,jordan2008,jordan2012}, or as a pulsational
reverse detonation \citep[PRD;][]{bravo2009a,bravo2009b}.  All of these
explosion mechanisms operate for near-$M_{\text{Ch}}$ CO white dwarfs
in single-degenerate systems.

The thermonuclear explosion of a sub-$M_{\text{Ch}}$ CO
white dwarf accreting helium can proceed as a double detonation
\citep{nomoto1982a,nomoto1982b,woosley1986,livne1990,fink2007,fink2010,woosley2011}.
In a double detonation, the compression of a helium envelope resulting
from accretion can spark a detonation in the helium envelope that produces
intermediate-mass elements.  That first detonation creates a shock wave
in the bulk of the white dwarf that subsequently sparks a detonation in
the interior and produces iron-peak elements.  The sub-$M_{\text{Ch}}$
CO white dwarf can accrete the helium shell necessary for a double
detonation from the helium core remaining after its companion star
loses its hydrogen envelope.  Accretion from a helium white dwarf
companion is also possible, either stably \citep[e.g.,][]{bildsten2007}
or dynamically \citep[e.g.,][]{webbink1984,guillochon2010,pakmor2013}.
Double detonations can therefore occur in both the single-degenerate
and double-degenerate scenarios.

Thermonuclear explosions of both $M_{\text{Ch}}$
and sub-$M_{\text{Ch}}$ CO white dwarfs can also be
attributed to detonations resulting from violent mergers
\citep[e.g.,][]{yoon2007,rosswog2009,pakmor2010,pakmor2011,pakmor2012}.
These violent mergers can occur in dense stellar environments or
as a result of secular eccentricity excitation in triple systems
\citep{thompson2011,kushnir2013}.  Thermonuclear explosions of this type
only occur through the double-degenerate scenario.

Despite these advances, traditional observations of Type Ia supernovae
host galaxies, progenitors, light curves, and remnants have been unable
to accurately and precisely quantify the relative contributions of each
of the progenitor systems and explosion mechanisms described above to
the population of thermonuclear events observed as Type Ia supernovae
\citep[e.g.,][]{maoz2014}.  A more accurate and precise understanding
of the progenitor systems and explosion mechanisms responsible for Type
Ia supernovae would be beneficial for numerous areas of astrophysics
from cosmology to galaxy formation to galactic chemical evolution
\citep[e.g.,][]{phillips1993,riess1998,perlmutter1999,kobayashi2020a,kobayashi2020b}.
In this article, we use published nucleosynthetic yields for thermonuclear
explosions that are prolific producers of iron to identify a region of
elemental abundance space consistent with those yields but rarely observed
in metal-poor stars.  We label stars in this elemental abundance space
``iron-rich metal-poor'' or IRMP stars and identify three such stars:
BD+80 245, HE 0533--5340, and SMSS J034249.53--284216.0.  We use a
state-of-the-art methodology to homogeneously infer elemental abundances
for these stars and compare those abundances to published nucleosynthetic
yields to identify the progenitor systems and/or explosion mechanism most
consistent with those abundances.  We describe in Section \ref{sample}
the construction of our sample.  We then infer stellar parameters for
our sample based on high-resolution optical spectra and all available
astrometric and photometric data in Section \ref{stellar_prop}.  We derive
the individual elemental abundances in Section \ref{elem_abund}.
We identify the progenitor system and/or explosion mechanism most
consistent with our elemental abundances the implications of those
results in Section \ref{discussion}.  We conclude by summarizing our
findings in Section \ref{conclusion}.

\section{Sample Definition}\label{sample}

\begin{figure*}
\plotone{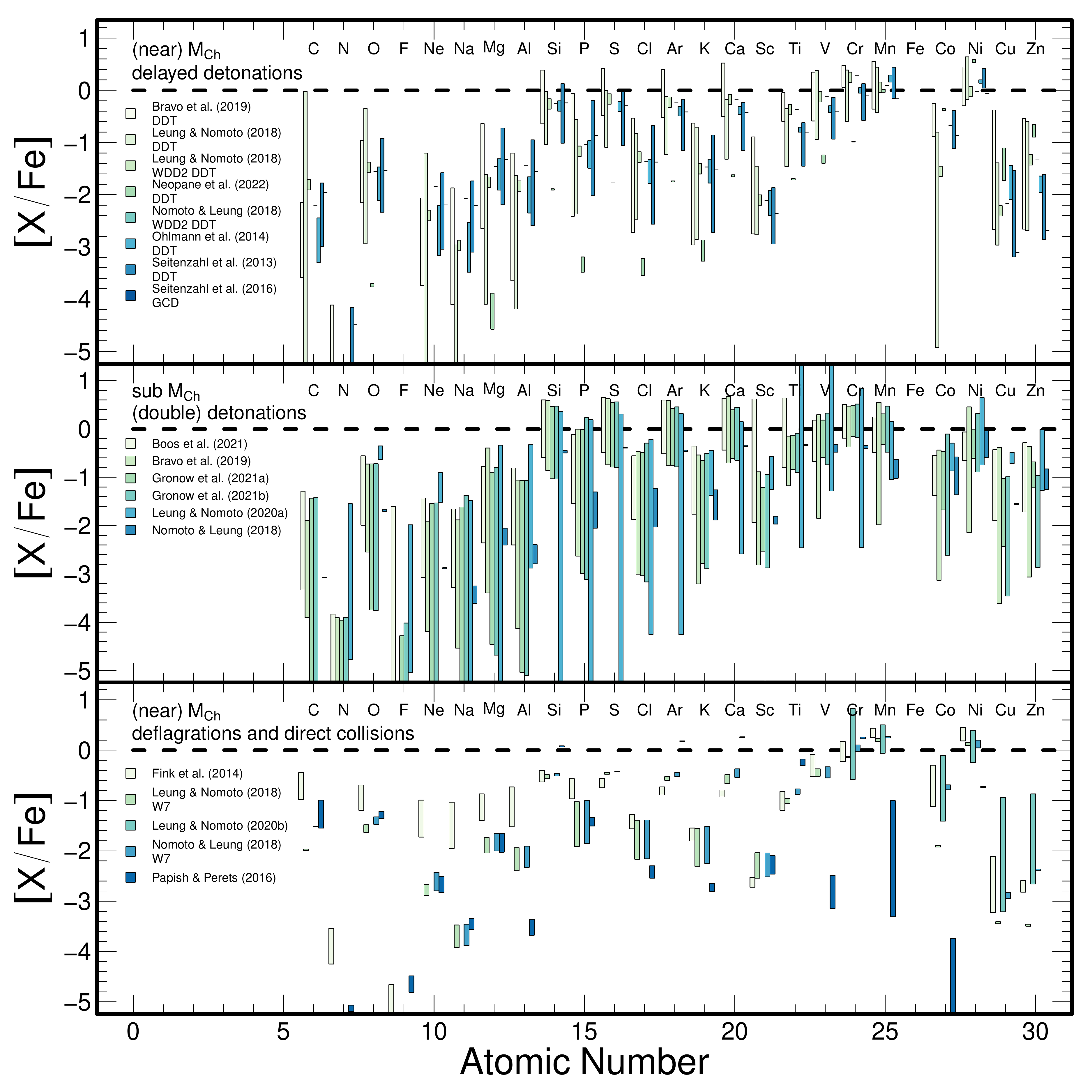} 
\caption{Predicted stable nucleosynthetic yields for thermonuclear
explosions that produce $M_{\text{Fe}} \geq 0.1~M_{\odot}$ with
varying progenitor systems and explosion mechanisms.  Each rectangle
represents the full range of predicted elemental abundances after
marginalizing over the parameters that define each progenitor system
and explosion mechanism.  Top: predicted stable elemental abundances
produced by delayed detonations.  Middle: predicted stable elemental
abundances produced by double detonations.  Bottom: predicted stable
elemental abundances produced by deflagrations and direct collisions.
Regardless of the progenitor system or explosion mechanism, relative
to iron the elemental abundances of carbon, nitrogen, oxygen, fluorine,
neon, sodium, magnesium, aluminum, chlorine, potassium, cobalt, copper,
and zinc are always sub-solar.  From this list, we use in our iron-rich
metal-poor star selection sodium, magnesium, and cobalt because those
abundances are usually straightforward to infer in the photospheres of
metal-poor giants.\label{yields}}
\end{figure*}

We seek to identify metal-poor stars with $[\text{Fe/H}] < -1$ formed
in regions dominated by thermonuclear nucleosynthesis.  We first
searched the literature for tables of stable nucleosynthetic yields
produced by thermonuclear explosions from any progenitor system and
explosion mechanism.\footnote{The yields in our compilation come
from \citet{seitenzahl2013,seitenzahl2016}, \citet{fink2014},
\citet{ohlmann2014}, \citet{kromer2015}, \citet{papish2016},
\citet{leung2018,leung2020a,leung2020b}, \citet{nomoto2018},
\citet{bravo2019}, \citet{boos2021}, \citet{gronow2021a,gronow2021b},
and \citet{neopane2022}.}  Because models of galactic chemical
evolution require thermonuclear explosions to be prolific
producers of iron, we only use models that produce $M_{\text{Fe}}
\geq 0.1~M_{\odot}$ to define the abundance space dominated by
thermonuclear nucleosynthesis.  We transform yields in mass to yields
in abundance ratio $[\text{X/Fe}]$ assuming \citet{asplund2021} solar
abundances and plot the results in Figure \ref{yields}.  While in
our yield compilation the variance in stable nucleosynthesis can be
large, there are commonalities.  We find that thermonuclear explosion
models that yield $M_{\text{Fe}} \geq 0.1~M_{\odot}$ always produce
$[\text{C,N,O,F,Ne,Na,Mg,Al,Cl,K,Co,Cu,Zn/Fe}] < 0$.  Carbon can be
synthesized in asymptotic giant branch (AGB) stars, transferred to binary
companions, and then transmuted to nitrogen.  An IRMP star definition
that depends on carbon and nitrogen could therefore fail to identify
stars formed in regions dominated by thermonuclear nucleosynthesis
subsequently impacted by mass transfer from an AGB star companion.
We therefore propose a useful working definition for IRMP stars robust to
mass transfer $[\text{O,F,Ne,Na,Mg,Al,Cl,K,Co,Cu,Zn/Fe}] < 0$.  Of these
elements, sodium, magnesium, and cobalt are most readily observable in the
photospheres of metal-poor giants and we focus our attention on stars with
published $[\text{Na,Mg,Co/Fe}] < 0$.  The requirement $[\text{Na,Mg/Fe}]
< 0$ tends to exclude stars enriched by ordinary core-collapse supernovae,
while the requirement $[\text{Co/Fe}] < 0$ should exclude stars enriched
by energetic explosions like hypernovae \citep[e.g.,][]{kobayashi2020a}.

We next searched the Stellar Abundances for Galactic Archaeology
(SAGA) database \citep{suda2008,suda2011,suda2017,yamada2013} for
giant stars with $[\text{Fe/H}] < -1$ and $[\text{Na,Mg,Co/Fe}] < 0$.
We identified four candidate iron-rich metal-poor giants: BD+80 0245,
HD 6833, HE 0533--5340, and SMSS J034249.53--284216.0.  The photospheric
stellar parameters and elemental abundances recorded in SAGA have been
inhomgeneously derived, and that makes it difficult to draw robust
conclusions from those archival data alone.  Consequently, we chose to
homogeneously analyze archival high-resolution, high signal-to-noise ratio
(S/N) optical spectra ourselves using the state-of-the-art methodology
described in \citet{reggiani2020,reggiani2021,reggiani2022b,reggiani2022a}
and Sections \ref{stellar_prop} and \ref{elem_abund}.

The uncommon elemental abundances of BD+80 245 were first noted by
\citet{carney1997}.  Its elemental abundances have been inferred many
times since, most recently by \citet{roederer2014} based on a spectrum
collected with the Tull Spectograph on the 2.7 m Harlan J.\ Smith
Telescope at McDonald Observatory \citep{tull1995}.  Ian Roederer kindly
provided to us that reduced, continuum-normalized spectrum of BD+80 245.
The discovery of HE 0533--5340 was reported in \citet{cohen2013} based
on a spectrum collected with the Magellan Inamori Kyocera Echelle
(MIKE) spectrograph on the Magellan Clay Telescope at Las Campanas
Observatory \citep{bernstein2003,shectman2003}.  Ian Thompson kindly
provided to us that reduced spectrum of HE 0533--5340.  The discovery
of SMSS J034249.53--284216.0 was reported in \citet{jacobson2015} based
on a Magellan/MIKE spectrum.  Heather Jacobson kindly provided to us
that reduced spectrum of SMSS J034249.53--284216.0.  The spectra of
both HE 0533--5340 and SMSS J034249.53--284216.0 were reduced using the
\texttt{CarPy}\footnote{\url{http://code.obs.carnegiescience.edu/mike}}
software package \citep{kelson2000,kelson2003,kelson2014}, and
we continuum normalized those spectra using Spectroscopy Made Harder
\citep[\texttt{smhr};][]{casey2014}\footnote{\url{https://github.com/andycasey/smhr/tree/py38-mpl313}}.
We report in Table \ref{obs_log} the properties of the spectra we analyze
as part of the study described in this article.

The importance of homogeneous analyses using
the state-of-the-art methodology described in
\citet{reggiani2020,reggiani2021,reggiani2022b,reggiani2022a} for the
realization of precise and accurate stellar parameters and elemental
abundances is illustrated by the case of HD 6833.  The elemental
abundances of HD 6833 have been reported by many authors, and we analyzed
an archival spectrum collected with High Resolution Echelle Spectrometer
(HIRES) on the Keck I Telescope at the Maunakea Observatories that
we downloaded from the Keck Observatory Archive \citep{vogt1994}.
We subsequently found that the priority photospheric stellar parameters
recorded in SAGA for HD 6833 from \citet{mishenina2017} were inaccurate,
and we dropped it from our sample because it does not meet our criteria
for IRMP stars.

\begin{deluxetable*}{lccccccccc}
\tabletypesize{\scriptsize}
\tablecaption{Observation Log\label{obs_log}}
\tablewidth{0pt}
\tablehead{
\colhead{Designation} &
\colhead{R.A.} &
\colhead{Decl.} & 
\colhead{UT Date} &
\colhead{Telescope/} &
\colhead{Exposure} &
\colhead{S/N} &
\colhead{S/N}\\
\colhead{} &
\colhead{(h:m:s)} &
\colhead{(d:m:s)} &
\colhead{} &
\colhead{Instrument} &
\colhead{Time (s)} &
\colhead{$4500~\rm{\AA}$} &
\colhead{$6500~\rm{\AA}$}
}
\startdata
BD+80 245                & 08:11:06.24 & +79:54:29.56 & 2009 May 06 & McDonald-Smith/Tull & 7200 & $70$ & $115$\\ 
HE 0533--5340             & 05:34:54.10 & $-$53:38:24.00 & 2003 Nov 01 & Magellan/MIKE & 9800 & $60$ & $100$\\  
SMSS J034249.53--284216.0 & 03:42:49.53 & $-$28:42:15.99& 2013 Jan 07 & Magellan/MIKE & 2000 & $45$ & $100$
\enddata
\end{deluxetable*}

\section{Stellar Properties}\label{stellar_prop}

\subsection{Stellar Orbits}\label{orbits}

We calculate Galactic orbits for the stars in our sample using
\texttt{Gala}\footnote{\url{http://gala.adrian.pw/en/latest/}}
\citep{gala2017,price_whelan2020}.  We first sample 1,000 Monte Carlo
realizations from the Gaia Data Release 3 (DR3) astrometric solutions
for each star using the distance posterior described below, taking full
account of the covariances between position, parallax, and proper motion
\citep{gaia2021,fabricius2021,lindegren2021a,lindegren2021b,torra2021}.
We used the radial velocities we measured from the spectra described
in Section \ref{sample} and assume no covariance between our measured
radial velocity and the Gaia DR3 astrometric solution.  We use
each Monte Carlo realization as an orbital initial condition and
integrate an orbit forward in time for 10 Gyr in a Milky Way-like
potential.  We adopted the \texttt{MWPotential2014} described by
\citet{bovy2015}.  In that model, the bulge is parameterized as a
power-law density profile that is exponentially cut-off at 1.9 kpc
with a power-law exponent of $-1.8$.  The disk is represented by a
Miyamoto--Nagai potential with a radial scale length of 3 kpc and a
vertical scale height of 280 pc \citep{miyamoto1975}.  The halo is
modeled as a Navarro--Frenk--White halo with a scale length of 16 kpc
\citep{navarro1996}.  We set the solar distance to the Galactic center to
$R_{0} = 8.122$ kpc, the circular velocity at the Sun to $V_{0} = 238$
km s$^{-1}$, the height of the Sun above the plane to $z_{0} = 25$ pc,
and the solar motion with the respect to the local standard of rest to
($U_{\odot}$, $V_{\odot}$, $W_{\odot}$) = (10.0, 11.0, 7.0) km s$^{-1}$
\citep{juric2008,blandhawthorn2016,gravity2018}.  We report the resulting
orbital parameters in Table \ref{stellar_prop_table}.  All three stars
are on ordinary inner halo-like orbits with no obvious evidence that
were formed in accreted and tidally disrupted dwarf galaxies.

\subsection{Fundamental and Photospheric Stellar Parameters}

We derive photospheric and fundamental stellar parameters
for the stars in our sample using the algorithm described in
\citet{reggiani2020,reggiani2021,reggiani2022b,reggiani2022a} that
makes use of both the classical spectroscopy-only approach\footnote{The
classical spectroscopy-only approach to photospheric stellar parameter
estimation involves simultaneously minimizing the difference between
\ion{Fe}{1} \& \ion{Fe}{2}-based abundances as well as their dependencies
on excitation potential and reduced equivalent width.} and isochrones
to infer accurate, precise, and self-consistent photospheric stellar
parameters.  Isochrones are especially useful for effective temperature
$T_{\text{eff}}$ inferences in this case, as high-quality multiwavelength
photometric data from the ultraviolet to the mid infrared are available
from the Galaxy Evolution Explorer (GALEX), the SkyMapper Southern Survey
(SMSS), Gaia, the Two-micron All-sky Survey (2MASS), and the Wide-field
Infrared Explorer (WISE). Similarly, the Gaia DR3 parallax-based distances
to the stars in our sample make the calculation of surface gravity
$\log{g}$ via isochrones straightforward.  With both $T_{\text{eff}}$ and
$\log{g}$ available via isochrones, the equivalent widths of iron lines
can be used to self-consistently determine metallicity $[\text{Fe/H}]$
and microturbulence $\xi$ by minimizing the dependence of individual
line-based iron abundance inferences on reduced equivalent width.
The microturbulence values inferred in this way can then be confirmed
using empirical relations \citep[e.g.,][]{kirby2011a}.

The inputs to our photospheric and fundamental stellar parameter
inference include the equivalent widths of \ion{Fe}{1} and
\ion{Fe}{2} atomic absorption lines, multiwavelength photometry, a
Gaia parallax, and an extinction estimate.  Using atomic absorption
line data from \cite{ji2020} based on the \texttt{linemake}
code\footnote{\url{https://github.com/vmplacco/linemake}}
\citep{sneden2009,sneden2016,placco2021} maintained by Vinicius
Placco and Ian Roederer, we first measure equivalent widths of atomic
absorption lines by fitting Gaussian profiles with \texttt{smhr} to a
continuum-normalized spectrum.  We gather
\begin{enumerate}
\item
NUV photometry and their uncertainties from GALEX \citep{bianchi2017};
\item
$u$, $v$, $g$, $r$, $i$, and $z$ photometry and their uncertainties from
SMSS DR2 \citep{onken2019};
\item
$G$ photometry and their uncertainties from Gaia DR2
\citep{gaia2016,gaia2018,arenou2018,evans2018,hambly2018,riello2018};
\item
$J$, $H$, and $K_{\text{s}}$ photometry and their uncertainties from
the 2MASS Point Source Catalog \citep{skrutskie2006}; and
\item
$W1$ and $W2$ photometry and their uncertainties from the WISE
AllWISE data release \citep{wright2010,mainzer2011}.
\end{enumerate}
We use Gaia DR3 parallaxes and their uncertainties
\citep{gaia2021,fabricius2021,lindegren2021a,lindegren2021b,torra2021}
as well as extinction $A_V$ inferences.  For BD+80 245,
HE 0533--5340, and SMSS J034249.53--284216.0 we take $A_V$
respectively from the three-dimensional (3D) Stilism reddening
maps \citep{lallement2014,lallement2018,capitanio2017}, the
\citet{schlegel1998} dust map as updated by \citet{schlafly2011}, and
the 3D Bayestar19 dust map \citep{green2019}.

We assume \citet{asplund2021} solar abundances and use these inputs
to infer photospheric and fundamental stellar parameters using the
following steps.
\begin{enumerate}
\item
We use 1D plane-parallel solar-composition ATLAS9 model atmospheres
\citep{castelli2004}, the 2019 version of the \texttt{MOOG}
radiative transfer code \citep{sneden1973}, and the \texttt{q$^2$}
\texttt{MOOG} wrapper\footnote{\url{https://github.com/astroChasqui/q2}}
\citep{ramirez2014} to derive an initial set of photospheric stellar
parameters $T_{\text{eff}}$, $\log{g}$, $[\text{Fe/H}]$, and $\xi$
using the classical spectroscopy-only approach.
\item
We then use the \texttt{isochrones}
package\footnote{\url{https://github.com/timothydmorton/isochrones}}
\citep{morton2015} to fit the MESA Isochrones and Stellar Tracks
\cite[MIST;][]{dotter2016,choi2016,paxton2011,paxton2013,paxton2015,paxton2018,paxton2019,jermyn2022}
library to our photospheric stellar parameters as well as our
input multiwavelength photometry, parallax, and extinction data using
\texttt{MultiNest}\footnote{\url{https://ccpforge.cse.rl.ac.uk/gf/project/multinest/}}
\citep{feroz2008,feroz2009,feroz2019} via \texttt{PyMultinest}
\citep{buchner2014}.  We restricted the MIST library to extinctions
between $A_{V}=0$ mag and the maximum suggested extinction for a
particular star plus 0.1 mag.  This produces a new set of photospheric
and fundamental stellar parameter posteriors that are both self-consistent
and physically consistent with stellar evolution.
\item
We next impose the posterior-median $T_{\text{eff}}$ and $\log{g}$
inferred in step 2 on our grid of model atmospheres and minimize the
dependence of individual line-based iron abundance inferences on reduced
equivalent width to derive model atmosphere $[\text{Fe/H}]_{\text{atmo}}$
and a new set of $\xi$ values consistent with our measured \ion{Fe}{1}
\& \ion{Fe}{2} equivalent widths and our \texttt{isochrones} inferred
$T_{\text{eff}}$ \& $\log{g}$.
\item
We then use the model atmosphere selected in step 3 to
calculate $[\text{Fe/H}]$ as the average of all $n_{\text{Fe}}
= n_{\text{\ion{Fe}{1}}} + n_{\text{\ion{Fe}{2}}}$ equivalent
width-based iron abundance inferences for individual \ion{Fe}{1} \&
\ion{Fe}{2} lines.  We take the uncertainty of our $[\text{Fe/H}]$
inference $\sigma_{[\text{Fe/H}]}$ as the standard deviation of the
individual line-based abundance inferences $\sigma_{[\text{Fe/H}]}'$
divided by $\sqrt{n_{\text{Fe}}}$.
\item
We next check if the $[\text{Fe/H}]$ inferred in step 4 agrees to two
decimal places with $[\text{Fe/H}]_{\text{atmo}}$.  If so, we proceed
to step 6.  If not, we replace $[\text{Fe/H}]_{\text{atmo}}$ with
$[\text{Fe/H}]$ and repeat steps 3 to 5 until agreement is achieved.
\item
We then repeat steps 2 to 5 until the metallicities inferred from
both the \texttt{isochrones} analysis and the reduced equivalent width
balance approach are consistent within their uncertainties (typically
a few iterations).
\end{enumerate}

We use a Monte Carlo simulation to derive the final values and
uncertainties in our adopted $[\text{Fe/H}]$ and $\xi$ values due to
the uncertainties in our adopted $T_{\text{eff}}$ and $\log{g}$.
\begin{enumerate}
\item
We randomly sample a self-consistent pair of $T_{\text{eff}}$ and
$\log{g}$ from our converged \texttt{isochrones} posteriors described
above and calculate the values of $[\text{Fe/H}]_{\text{atmo}}$ and
$\xi$ that produce the best reduced equivalent width balance given our
\ion{Fe}{1} \& \ion{Fe}{2} equivalent width measurements.
\item
We use the model atmosphere selected in step 1 to calculate the average
of all $n_{\text{Fe}} = n_{\text{\ion{Fe}{1}}} + n_{\text{\ion{Fe}{2}}}$
individual equivalent width-based iron abundance inferences and save
the resulting metallicity of each iteration.
\item
We repeat steps 1 and 2 200 times and adopt as our final photospheric
stellar parameters the (16,50,84) percentiles of the 200 self-consistent
sets of $T_{\text{eff}}$, $\log{g}$, $[\text{Fe/H}]$, and $\xi$ produced
in this way.
\end{enumerate}
We find good agreement between these photospheric stellar parameters
derived from our Monte Carlo simulation and those resulting from
a single iteration of reduced equivalent width balance using the
median $T_{\text{eff}}$ and $\log{g}$ from the posteriors produced
by our converged analysis.  We report our adopted photospheric and
fundamental stellar parameters in Table \ref{stellar_prop_table}. All
of the uncertainties quoted in Table \ref{stellar_prop_table} include
random uncertainties only.  That is, they are uncertainties derived
under the unlikely assumption that the MIST isochrone grid we use in
our analyses perfectly reproduces all stellar properties.

\begin{deluxetable*}{lccccc} 
\tabletypesize{\scriptsize}
\tablecaption{Stellar Properties and Adopted
Parameters\label{stellar_prop_table}}
\tablewidth{0pt} 
\tablehead{ 
\colhead{} & \colhead{} & \colhead{} & \colhead{} & \colhead{} & \colhead{}}
\startdata
Property & BD+80 0245 & HE 0533--5340 & SMSS J034249.53--284216.0 & Units \\
Gaia DR3 \texttt{source\_id} & 1139085117140997120 & 4768665767426450304 & 5080395975535412224 & \\
\hline
\textbf{Astrometric Properties} & & & \\
Gaia DR3 parallax $\pi$ & $4.312\pm0.012$ & $0.062\pm0.020$ & $0.095\pm0.016$ & mas \\
Gaia DR3 proper motion $\mu_{\alpha} \cos{\delta}$ & $136.699\pm0.0132$ & $6.025\pm0.026$ & $2.024\pm0.011$ & mas yr$^{-1}$ \\
Gaia DR3 proper motion $\mu_{\delta}$ & $-367.492\pm0.012$ & $2.756\pm0.031$ & $-1.058\pm0.015$ & mas yr$^{-1}$ \\
\hline 
\textbf{Photometric Properties} & & & \\ 
GALEX $NUV$   & $\cdots$ & $20.402\pm0.099$ & $\cdots$ & AB mag\\
SkyMapper $u$ & $\cdots$ & $16.926\pm0.041$ & $16.357\pm0.035$ & AB mag \\ 
SkyMapper $v$ & $\cdots$ & $16.444\pm0.019$ & $\cdots$ & AB mag \\ 
SkyMapper $g$ & $\cdots$ & $15.309\pm0.013$ & $14.645\pm0.011$ & AB mag \\ 
SkyMapper $r$ & $\cdots$ & $14.890\pm0.005$ & $14.270\pm0.008$ & AB mag \\ 
SkyMapper $i$ & $\cdots$ & $\cdots$ & $13.950\pm0.006$ & AB mag \\ 
SkyMapper $z$ & $\cdots$ & $14.428\pm0.008$ & $\cdots$ & AB mag \\ 
Gaia DR2 $G$ & $9.803\pm0.002$ & $14.860\pm0.002$ & $14.243\pm0.002$ & Vega mag \\
2MASS $J$ & $8.711\pm0.039$ & $13.416\pm0.028$ & $12.837\pm0.024$ & Vega mag \\ 
2MASS $H$ & $8.333\pm0.038$ & $12.894\pm0.024$ & $12.307\pm0.024$ & Vega mag \\ 
2MASS $K_{\text{s}}$ & $8.261\pm0.026$ & $12.811\pm0.029$ & $12.224\pm0.026$ & Vega mag \\ 
WISE W1 & $8.210\pm0.023$ & $12.761\pm0.023$ & $12.168\pm0.023$ & Vega mag \\ 
WISE W2 & $8.244\pm0.021$ & $12.754\pm0.022$ & $12.182\pm0.022$ & Vega mag \\ 
\hline
\textbf{Stellar Properties} & & & \\ 
Luminosity $L_{\ast}$ & $5\pm1$ & $143^{+2}_{-5}$ & $114^{+5}_{-7}$ & L$_{\odot}$ \\ 
Radius $R_{\ast}$ & $2\pm1$ & $15\pm1$ & $14\pm1$ & R$_{\odot}$ \\ 
Distance $d_{\text{iso}}$ & $0.2\pm0.1$ & $11.2^{+0.1}_{-0.2}$ & $7.8^{+0.2}_{-0.3}$ & kpc \\ 
Mass $M_{\odot}$ & $0.79^{+0.02}_{-0.01}$ & $0.82^{+0.03}_{-0.02}$ & $0.86^{+0.09}_{-0.06}$ & M$_{\odot}$ \\ 
Age $\tau$ & $12.3^{+0.7}_{-0.9}$ & $12.3^{+0.9}_{-1.4}$ & $9.2^{+2.6}_{-2.5}$ & Gyr \\ 
Extinction $A_{V}$ & $0.090^{+0.038}_{-0.028}$ & $0.219^{+0.015}_{-0.014}$ & $0.138^{+0.029}_{-0.027}$ & mag \\ 
Effective temperature $T_{\text{eff}}$ & $5696^{+47}_{-36}$ & $5031^{+11}_{-9}$ & $5017^{+26}_{-23}$ & K \\ 
Surface gravity $\log{g}$ & $3.57\pm0.01$ & $1.94\pm0.02$ & $2.08\pm0.04$ & cm s$^{-2}$ \\ 
Metallicity $[\text{Fe/H}]$ & $-1.73\pm0.11$ & $-2.44\pm0.15$ & $-1.97\pm0.17$ &  \\ 
Microturbulence $\xi$ & $1.31\pm0.10$ & $1.68\pm0.10$ & $1.65\pm0.10$ & km s$^{-1}$ \\
\hline
\textbf{Orbital Properties} & & & \\ 
Radial velocity $v_{r}$ & $5.0\pm0.2$ & $147.2\pm1.5$ & $156.7\pm3.5$ & km s$^{-1}$ \\
Total Galactic velocity $v$ & $185.5^{+114.5}_{-62.0}$ & $211.6^{+28.0}_{-24.3}$ & $196.4^{+100.4}_{-56.6}$ & km s$^{-1}$ \\
Pericenter $R_{\text{peri}}$ & $5.28^{+0.04}_{-0.04}$ & $11.61^{+3.19}_{-5.71}$ & $5.43^{+0.23}_{-0.29}$ & kpc \\
Apocenter $R_{\text{apo}}$ & $15.83^{+0.21}_{-0.20}$ & $15.66^{+14.50}_{-3.41}$ & $13.56^{+0.95}_{-0.91}$ & kpc \\
Eccentricity $e$ & $0.5\pm0.1$ & $0.26^{+0.19}_{-0.10}$ & $0.43^{+0.05}_{-0.05}$ & \\
Maximum distance from Galactic plane $z_{\text{max}}$ & $9.48^{+0.08}_{-0.08}$ & $12.61^{+8.18}_{-4.83}$ & $5.25^{+0.34}_{-0.35}$ & kpc \\
Angular momentum $L_{z}$ & $0.7\pm0.1$ & $-1.1\pm0.10$ & $-1.4\pm0.1$ & kpc km s$^{-1}$ \\
Specific orbital energy $E_{\text{tot}}$ & $-1.2\pm0.1$ & $-1.1\pm0.1$ & $-1.3\pm0.1$ & $10^{5}~\text{km}^{2}$ s$^{-2}$ \\
\enddata
\tablecomments{We report random uncertainties derived under the unlikely
assumption that the MIST isochrone grid perfectly reproduces all stellar
properties.  There are almost certainly larger systematic uncertainties
that we have not investigated, though the excellent agreement between our
analysis and previous results for the three stars in our sample indicate
that any systematic uncertainties in our analysis cannot be too large.}
\end{deluxetable*} 

To evaluate the impact of any possible systematic uncertainties resulting
from our analysis, we compare the photospheric stellar parameters we
infer for the stars in our sample with those reported by other groups
for the same stars.  Photospheric stellar parameters for BD+80 245 based
on high-resolution optical spectra using modern analysis techniques have
been presented by many authors.
\begin{enumerate}
\item
\citet{carney1997} found $T_{\text{eff}} = 5400 \pm 100$ K, $\log{g} =
3.2 \pm 0.14$, $[\text{Fe/H}] = -1.86 \pm 0.05$, and $\xi = 1.5 \pm 0.2$
km s$^{-1}$;
\item
\citet{fulbright2000} found $T_{\text{eff}} = 5225 \pm 40$ K, $\log{g} =
3.0 \pm 0.06$, $[\text{Fe/H}] = -1.90 \pm 0.08$, and $\xi = 1.35\pm0.11$
km s$^{-1}$;
\item
\citet{stevens2002} found $T_{\text{eff}} = 5569 \pm 121$ K, $\log{g}
= 3.47 \pm 0.46$, $[\text{Fe/H}]= -1.76 \pm 0.09$, and $\xi = 1.56 \pm
0.09$ km s$^{-1}$;
\item
\citet{ivans2003} found $T_{\text{eff}} = 5225$ K, $\log{g} = 3.00$,
$[\text{Fe/H}] = -2.07$, and $\xi = 1.25$ km s$^{-1}$;
\item
\citet{zhang2005} found $T_{\text{eff}} = 5446 \pm 100$ K, $\log{g} =
3.31 \pm 0.20$, $[\text{Fe/H}]=-1.72 \pm 0.10$, and $\xi = 1.90 \pm 0.50$
km s$^{-1}$; and
\item
\citet{roederer2014} found $T_{\text{eff}} = 5360 \pm 34$ K, $\log{g}
= 3.15 \pm 0.22$, $[\text{M/H}] = -2.01 \pm 0.07$, and $\xi = 1.20 \pm
0.06$ km s$^{-1}$.
\end{enumerate}
HE 0533--5340 was initially studied by \citet{cohen2013} in the paper
announcing its discovery, and those authors found $T_{\text{eff}} =
4937 \pm 100$ K, $\log{g} = 1.80 \pm 0.25$, $[\text{Fe/H}] = -2.67 \pm
0.20$, and $\xi = 2.1 \pm 0.2$ km s$^{-1}$.  SMSS J034249.53--284216.0
was initially studied by \citet{jacobson2015} in the paper announcing
its discovery, and those authors found $T_{\text{eff}} = 4828 \pm 100$
K, $\log{g} = 1.91 \pm 0.3$, $[\text{Fe/H}] = -2.33 \pm 0.14$, and
$\xi = 1.95 \pm 0.2$ km s$^{-1}$.  In general, our photospheric
stellar parameters are consistent with those previously inferred for
the same stars.  These previous analyses mostly used the classical
spectroscopy-only approach in the pre-Gaia era though, and we attribute
any differences between our and previous inferences to the well-known
tendency of spectroscopy-only inferences to obtain cooler temperatures
and lower surface gravities than photometry- and parallax-informed
methods \citep[e.g.,][]{korn2003,frebel2013,mucciarelli2020}.

As an additional check, we infer $T_{\text{eff}}$ using the
\texttt{colte} code\footnote{\url{https://github.com/casaluca/colte}}
\citep{casagrande2021} that estimates $T_{\text{eff}}$ using a combination
of color--$T_{\text{eff}}$ relations obtained by implementing the InfraRed
Flux Method for Gaia and 2MASS photometry.  As required by \texttt{colte},
we used Gaia DR3 $G$, $G_{\text{BP}}$, and $G_{\text{RP}}$ plus 2MASS $J$,
$H$, and $K_{\text{s}}$ photometry as input. Using this approach,
we find $T_{\text{eff}} = 5601 \pm 61$ K for BD+80 0245, $T_{\text{eff}}
= 5003 \pm 59$ K for HE 0533--5340, and $T_{\text{eff}} = 4943 \pm 53$
K for SMSS J034249.53--285216.0.  These IRFM-based temperatures are
consistent with our adopted values.  We are therefore confident than any
systematic uncertainties arising from our methodology should be small.

\section{Elemental Abundances}\label{elem_abund}

To infer the elemental abundances of several $\alpha$, light odd-$Z$,
iron-peak, and neutron-capture elements we first measure the equivalent
widths of atomic absorption lines of \ion{O}{1}, \ion{Na}{1}, \ion{Mg}{1},
\ion{Al}{1}, \ion{Si}{1}, \ion{Ca}{1}, \ion{Sc}{2}, \ion{Ti}{1},
\ion{Ti}{2}, \ion{Cr}{1}, \ion{Cr}{2}, \ion{Mn}{1}, \ion{Fe}{1},
\ion{Fe}{2}, \ion{Ni}{1}, \ion{Zn}{1}, \ion{Sr}{2}, \ion{Y}{2}, and
\ion{Ba}{2} in our continuum-normalized spectra by fitting Gaussian
or Voigt profiles as appropriate with \texttt{smhr}.  We use atomic
absorption line data from \cite{ji2020} based on the \texttt{linemake}
code\footnote{\url{https://github.com/vmplacco/linemake}}
\citep{sneden2009,sneden2016,placco2021}.  We measure an equivalent width
for every absorption line in our line list that could be recognized,
taking into consideration the quality of the spectrum in the vicinity of a
line and the availability of alternative transitions of the same species.
We assume \citet{asplund2021} solar abundances and local thermodynamic
equilibrium (LTE) and use the 1D plane-parallel solar-composition
ATLAS9 model atmospheres and the 2019 version of \texttt{MOOG} to
infer elemental abundances based on each equivalent width measurement.
We report our adopted atomic data, equivalent width measurements, and
individual line-based abundance inferences in Table \ref{measured_ews}.

\begin{deluxetable*}{llccccc}
\tablecaption{Atomic Data, Equivalent-width Measurements, and
Individual-line Abundance Inferences\label{measured_ews}}
\tablewidth{0pt}
\tablehead{
\colhead{Star} & \colhead{Wavelength} & \colhead{Species} &
\colhead{Excitation Potential} & \colhead{log($gf$)} &
\colhead{EW} & \colhead{$\log_\epsilon(\rm{X})$} \\
 & \colhead{(\AA)} &  & \colhead{(eV)} & & (m\AA) & }
\startdata
BD+80 0245 & $5889.951$ & \ion{Na}{1} & $0.000$ & $0.108$ & $160.71$ & $4.402$\\ 
HE 0533--5340 & $5889.951$ & \ion{Na}{1} & $0.000$ & $0.108$ & $110.31$ & $3.174$\\
SMSS J034249.53--284216.0 & $5889.951$ & \ion{Na}{1} & $0.000$ & $0.108$ & $154.12$ & $3.865$\\
BD+80 0245 & $3986.753$ & \ion{Mg}{1} & $4.346$ & $-1.060$ & $29.85$ & $5.659$\\ 
BD+80 0245 & $4057.505$ & \ion{Mg}{1} & $4.346$ & $-0.900$ & $36.30$ & $5.622$\\ 
BD+80 0245 & $4571.096$ & \ion{Mg}{1} & $0.000$ & $-5.623$ & $20.28$ & $5.732$\\ 
HE 0533--5340 & $4057.505$ & \ion{Mg}{1} & $4.346$ & $-0.900$ & $15.70$ & $4.877$\\
HE 0533--5340 & $4167.271$ & \ion{Mg}{1} & $4.346$ & $-0.745$ & $27.07$ & $5.011$\\
HE 0533--5340 & $4571.096$ & \ion{Mg}{1} & $0.000$ & $-5.623$ & $16.56$ & $4.765$\\
SMSS J034249.53--284216.0 & $3986.753$ & \ion{Mg}{1} & $4.346$ & $-1.060$ & $17.88$ & $5.092$\\
SMSS J034249.53--284216.0 & $4057.505$ & \ion{Mg}{1} & $4.346$ & $-0.900$ & $41.11$ & $5.419$\\
SMSS J034249.53--284216.0 & $4167.271$ & \ion{Mg}{1} & $4.346$ & $-0.745$ & $53.94$ & $5.473$\\
\enddata
\tablecomments{This table is published in its entirety in the
machine-readable format.  A portion is shown here for guidance regarding
its form and content.}
\end{deluxetable*}

We use spectral synthesis to infer the abundances of elements
that are difficult or impossible to measure in our spectra using
equivalent widths.  Cobalt lines in our spectra are too weak for the
equivalent width approach, so we inferred cobalt abundances using up
to six \ion{Co}{1} lines at 3894, 3995, 4020, 4110, 4118, and 4121~\AA.
We also use spectral synthesis to infer europium abundances using up to
three \ion{Eu}{2} lines at 4129, 4435, and 4522~\AA.  We account for the
effects of hyperfine and/or isotopic splitting on lines of \ion{Sc}{2},
\ion{Mn}{1}, \ion{Co}{1}, \ion{Y}{2}, \ion{Ba}{2}, and \ion{Eu}{2} using
data from Kurucz\footnote{\url{http://kurucz.harvard.edu/linelists.html}
} supplemented by data from \citet{mcwilliam1998} and \citet{klose2002}
for barium.  We present our adopted mean elemental abundances
and uncertainties in Table \ref{elem_abundances}.  We define the
uncertainty in the abundance ratio $\sigma_{[\text{X/H}]}$ as the
standard deviation of the individual line-based abundance inferences
$\sigma_{[\text{X/H}]}'$ divided by $\sqrt{n_{\text{X}}}$.  We define
the uncertainty $\sigma_{[\text{X/Fe}]}$ as the square root of the sum
of squares of $\sigma_{[\text{X/H}]}$ and $\sigma_{[\text{Fe/H}]}$.

\begin{deluxetable*}{lcccccccccccc}
\tablecaption{Elemental Abundances\label{elem_abundances}}
\tablecolumns{13}
\tabletypesize{\small}
\tablewidth{0pt}
\tablehead{
\colhead{Species} &
\colhead{n} & \colhead{log($\epsilon_X$)} &
\colhead{[X/Fe]} & \colhead{$\sigma_{[\text{X/Fe}]}$} &
\colhead{n} & \colhead{log($\epsilon_X$)} &
\colhead{[X/Fe]} & \colhead{$\sigma_{[\text{X/Fe}]}$} &
\colhead{n} & \colhead{log($\epsilon_X$)} &
\colhead{[X/Fe]} & \colhead{$\sigma_{[\text{X/Fe}]}$} }
\startdata
\hline
 & \multicolumn{4}{c}{BD+80 0245} & \multicolumn{4}{c}{HE 0533--5340} & \multicolumn{4}{c}{SMSS J034249.53--284216.0} \\
 \ion{O}{1} & $2$ & $7.325$ & $0.365$ & $0.273$ & $\cdots$ & $\cdots$ & $\cdots$ & $\cdots$ & $\cdots$ & $\cdots$ & $\cdots$ & $\cdots$ \\
\ion{Na}{1} & $2$ & $4.438$ & $-0.052$ & $0.077$ & $2$ & $3.174$ & $-0.604$ & $0.045$ & $2$ & $3.788$ & $-0.463$ & $0.105$  \\ 
\ion{Na}{1}$_{\rm{NLTE}}$ & $2$ & $3.963$ & $-0.527$ & $\cdots$ & $2$ & $2.901$ & $-0.877$ & $\cdots$ & $2$ & $3.325$ & $-0.926$ & $\cdots$  \\ 
\ion{Mg}{1} & $6$ & $5.615$ & $-0.205$ & $0.059$ & $7$ & $4.855$ & $-0.253$ & $0.036$ & $7$ & $5.300$ & $-0.281$ & $0.069$  \\ 
\ion{Mg}{1}$_{\rm{NLTE}}$ & $1$ & $5.772$ & $-0.095$ & $\cdots$ & $\cdots$ & $\cdots$ & $\cdots$ & $\cdots$ & $1$ & $5.429$ & $-0.167$ & $\cdots$  \\ 
\ion{Al}{1} & $1$ & $3.163$ & $-1.537$ & $0.067$ & $1$ & $2.600$ & $-1.388$ & $0.049$ & $1$ & $3.066$ & $-1.395$ & $0.073$  \\ 
\ion{Si}{1} & $1$ & $5.759$ & $-0.021$ & $0.052$ & $1$ & $5.546$ & $0.478$ & $0.027$ & $1$ & $5.769$ & $0.228$ & $0.052$  \\ 
\ion{Si}{1}$_{\rm{NLTE}}$ & $1$ & $5.803$ & $-0.024$ & $\cdots$ & $1$ & $5.524$ & $0.426$ & $\cdots$ & $1$ & $5.738$ & $0.182$ & $\cdots$  \\ 
\ion{Ca}{1} & $23$ & $4.306$ & $-0.264$ & $0.046$ & $14$ & $3.753$ & $-0.105$ & $0.041$ & $18$ & $4.154$ & $-0.177$ & $0.068$  \\ 
\ion{Sc}{2} & $2$ & $0.859$ & $-0.551$ & $0.077$ & $1$ & $-0.014$ & $-0.712$ & $0.014$ & $3$ & $0.696$ & $-0.475$ & $0.225$  \\ 
\ion{Ti}{1} & $21$ & $3.201$ & $-0.039$ & $0.112$ & $14$ & $2.316$ & $-0.212$ & $0.041$ & $30$ & $2.883$ & $-0.118$ & $0.093$  \\ 
\ion{Ti}{2} & $29$ & $3.078$ & $-0.162$ & $0.051$ & $32$ & $2.530$ & $0.002$ & $0.055$ & $43$ & $2.934$ & $-0.067$ & $0.067$  \\ 
\ion{Cr}{1} & $16$ & $3.799$ & $-0.091$ & $0.060$ & $10$ & $2.890$ & $-0.288$ & $0.065$ & $14$ & $3.263$ & $-0.388$ & $0.058$  \\ 
\ion{Cr}{2} & $5$ & $3.821$ & $-0.069$ & $0.040$ & $2$ & $3.214$ & $0.036$ & $0.202$ & $5$ & $3.362$ & $-0.289$ & $0.042$  \\ 
\ion{Mn}{1} & $4$ & $3.466$ & $-0.224$ & $0.110$ & $1$ & $3.069$ & $0.091$ & $0.012$ & $2$ & $3.151$ & $-0.300$ & $0.048$  \\ 
\ion{Fe}{1} & $117$ & $5.739$ & $\cdots$ & $\cdots$  & $115$ & $5.026$ & $\cdots$ & $\cdots$  & $121$ & $5.480$ & $\cdots$ & $\cdots$   \\
\ion{Fe}{1}$_{\rm{NLTE}}$ & $117$ & $5.874$ & $\cdots$ & $0.092$  & $115$ & $5.053$ & $\cdots$ & $0.113$  & $121$ & $5.543$ & $\cdots$ & $0.148$   \\
\ion{Fe}{2} & $23$ & $5.682$ & $\cdots$ & $\cdots$ & $19$ & $4.970$ & $\cdots$ & $\cdots$ & $25$ & $5.542$ & $\cdots$ & $\cdots$  \\ 
\ion{Fe}{2}$_{\rm{NLTE}}$ & $23$ & $5.727$ & $\cdots$ & $0.115$ & $19$ & $4.998$ & $\cdots$ & $0.101$ & $25$ & $5.569$ & $\cdots$ & $0.160$  \\ 
\ion{Co}{1} & $5$ & $3.230$ & $-0.023$ & $0.100$ & $6$ & $2.250$ & $-0.273$ & $0.150$ & $6$ & $2.620$ & $-0.370$ & $0.150$  \\ 
\ion{Ni}{1} & $21$ & $4.367$ & $-0.103$ & $0.063$ & $11$ & $3.729$ & $-0.029$ & $0.125$ & $22$ & $4.099$ & $-0.132$ & $0.065$  \\ 
\ion{Zn}{1} & $2$ & $2.379$ & $-0.451$ & $0.035$ & $\cdots$ & $\cdots$ & $\cdots$ & $\cdots$ & $1$ & $2.182$ & $-0.409$ & $0.017$  \\ 
\ion{Sr}{2} & $2$ & $0.293$ & $-0.807$ & $0.228$ & $2$ & $-0.149$ & $-0.537$ & $0.097$ & $2$ & $0.575$ & $-0.286$ & $0.066$  \\ 
\ion{Y}{2} & $\cdots$ & $\cdots$ & $\cdots$ & $\cdots$ & $\cdots$ & $\cdots$ & $\cdots$ & $\cdots$ & $2$ & $-0.385$ & $-0.626$ & $0.212$  \\ 
\ion{Ba}{2} & $2$ & $-0.887$ & $-1.427$ & $0.057$ & $4$ & $-1.178$ & $-1.006$ & $0.052$ & $4$ & $0.231$ & $-0.070$ & $0.130$  \\ 
\ion{Ba}{2}$_{\rm{NLTE}}$ & $\cdots$ & $\cdots$ & $\cdots$ & $\cdots$ & $1$ & $-1.079$ & $-0.937$ & $\cdots$ & $3$ & $-0.02$ & $-0.336$ & $0.076$  \\ 
\ion{Eu}{2} & $1$ & $\leq-1.700$ & $\leq-0.533$ & $\cdots$ & $1$ & $\leq-2.410$ & $\leq-0.513$ & $\cdots$ & $1$ & $-1.140$ & $0.290$ & $0.200$
\enddata
\end{deluxetable*}

\subsection{$\alpha$ Elements}\label{alpha}

Oxygen, magnesium, silicon, calcium, and titanium are often referred to
as $\alpha$ elements.  Oxygen is produced in helium and neon burning
($^{16}$O) as well as the CNO tri-cycle in hydrogen shell burning
($^{17}$O) and through $\alpha$ capture by $^{14}$N during helium
shell burning ($^{18}$O) \citep{woosley1995}.  Magnesium, silicon, and
calcium are formed via similar nucleosynthetic channels.  Magnesium is
mainly synthesized by carbon burning in core-collapse supernovae, and
at solar metallicities thermonuclear events provide at least an order
of magnitude less.  Silicon is mostly a product of oxygen burning and is
itself the most abundant product of oxygen burning.  At solar metallicity
core-collapse and thermonuclear supernovae contribute to equally to
silicon production.  Calcium is the product of both hydrostatic and
explosive oxygen and silicon burning.  At solar metallicity it is
mostly produced in core-collapse supernovae.  Even though titanium
forms either in the $\alpha$-rich freeze-out of shock-decomposed
nuclei during core-collapse supernovae or in explosive $^{4}$He fusion
in the envelopes of CO white dwarfs during thermonuclear explosions
\citep[e.g.,][]{woosley1994,livne1995}, it is often considered alongside
the true $\alpha$ elements because of their similar chemical abundances
\citep{clayton2003}.  At solar metallicity both core-collapse and
thermonuclear explosions are important sources of titanium.   Magnesium,
silicon, calcium, and titanium abundances inferences based on the lines we
use are not strongly affected by departures from the assumptions of LTE.
Nevertheless, we correct magnesium and silicon abundances for departures
from the assumptions of LTE (i.e., non-LTE or NLTE) using data from
\citet{osorio2015} and \citet{amarsi2017}.

\begin{figure*}
\plotone{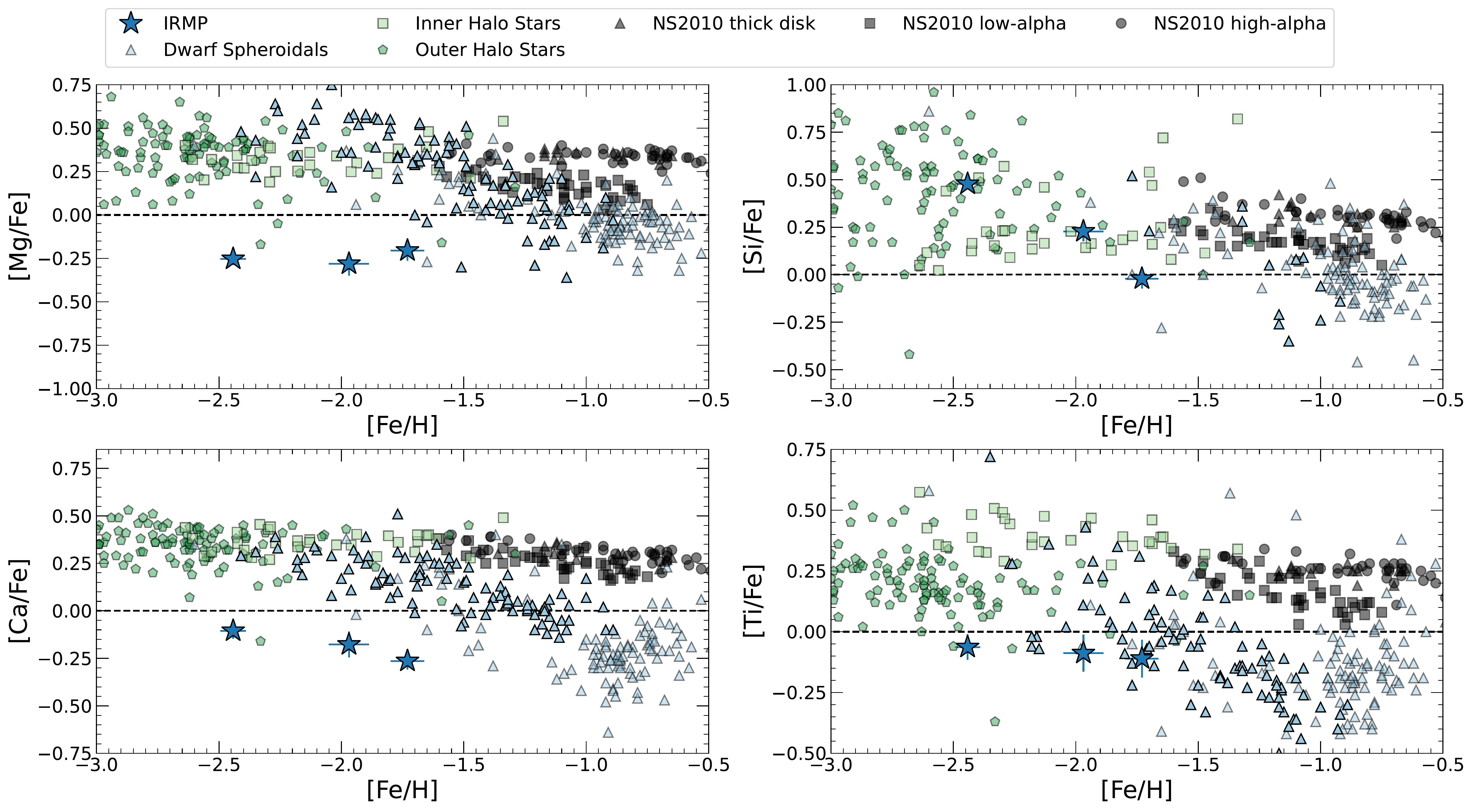}
\caption{Abundances of titanium and the $\alpha$-elements magnesium,
silicon, and calcium.  We plot as dark-blue stars the three iron-rich
metal-poor stars in our sample.   We plot as black-bordered light-green
squares inner halo metal-poor stars from \citet{reggiani2017} \&
Reggiani et al.\ (2023, in prep); as dark gray triangles, squares,
and circles thick disk, low-$\alpha$, and high-$\alpha$ stars from
\citet{nissen2010}; as black-boarded light-blue triangles stars from
the dwarf spheroidal galaxies Carina, Fornax, Sagittarius, and Sculptor
\citep{shetrone2003,geisler2005,monaco2005,letarte2010,hill2019,skuladottir2019};
and as dark-green pentagons outer halo metal-poor stars from
\citet{cayrel2004} \& \citet{jacobson2015}.  By construction, our
three IRMP stars have low $[\text{Mg/Fe}]$ abundances.  Their low
$[\text{Ca/Fe}]$ and $[\text{Ti/Fe}]$ abundances generally disfavor
spot- or belt-like double detonations with either especially thin or
especially thick helium envelopes.  Direct white dwarf collisions are
also disfavored.\label{alpha_fig}}
\end{figure*}

We report in Table \ref{elem_abundances} the magnesium, silicon, calcium,
and titanium abundances we infer for the three IRMP stars in our sample
and plot them in Figure \ref{alpha_fig} along with several comparison
samples.  While $[\text{Mg/Fe}] < 0$ was required for these three
IRMP stars by construction, they also have subsolar $[\text{Ca/Fe}]$
and $[\text{Ti/Fe}]$ abundances.  Subsolar $[\text{$\alpha$/Fe}]$
at $[\text{Fe/H}] \lesssim -1$ are observed in classical dwarf
spheroidal (dSph) galaxies, and Milky Way stars with this property
are often attributed to the accretion of now-disrupted dSph galaxies.
As is apparent in Figure \ref{alpha_fig} though, the $[\text{Mg/Fe}]$,
$[\text{Ca/Fe}]$, and $[\text{Ti/Fe}]$ abundances we observe in these
three IRMP stars are low even for stars at similar metallicities in
dSph galaxies.  As will argue in Section \ref{discussion}, while the
$\alpha$-element abundances of these three stars might be attributed to
formation in dSph galaxies the complete set of each IRMP star's elemental
abundances does not support this interpretation.

By themselves the subsolar abundances of $[\text{Ca/Fe}]$
and $[\text{Ti/Fe}]$ we observe generally disfavor thermonuclear
explosions produced by white dwarf collisions or spot- or belt-like
double detonations with either especially thin or especially thick
helium envelopes.  More specifically, subsolar $[\text{Ca/Fe}]$
abundances disfavor:
\begin{enumerate}
\item
double-detonation models from \citet{boos2021} with total masses
$M_{\text{tot}} \leq 0.9~M_{\odot}$ and thin helium shells
$\rho_{\text{base}} \leq 3 \times 10^{5}$ g cm$^{-3}$;
\item
$M_{\text{Ch}}$-mass DDT models from \citet{bravo2019} with low densities
ahead of the flame at the moment the DDT occurs $\rho_{\text{DDT}} \leq
1.6 \times 10^{7}$ g cm$^{-3}$ at all metallicities and independent of
the uncertain $^{12}$C + $^{16}$O reaction rate;
\item
sub-$M_{\text{Ch}}$ mass detonation models from \citet{bravo2019} with
white dwarf masses $M_{\text{WD}} \leq 0.97~M_{\odot}$ except for the
highest-metallicity $M_{\text{WD}} = 0.97~M_{\odot}$ model with $Z =
6.75 \times 10^{-2}$;
\item
double-detonation models from \citet{gronow2021a,gronow2021b} with initial
core masses $M_{\text{c\_ini}} \lesssim 0.8~M_{\odot}$ and helium shell
masses $M_{\text{c\_ini}} \lesssim 0.1~M_{\odot}$ as well as the model
with $M_{\text{c\_ini}} = 0.905~M_{\odot}$ and the lowest helium shell
masses $M_{\text{c\_ini}} = 0.026~M_{\odot}$; and
\item
white dwarf collisions models from \citet{papish2016}.
\end{enumerate}
Subsolar $[\text{Ti/Fe}]$ abundances disfavor:
\begin{enumerate}
\item
double-detonation models from \citet{boos2021} with total
masses $M_{\text{tot}} = 1.0~M_{\odot}$ and thick helium shells
$\rho_{\text{base}} \geq 6 \times 10^{5}$ g cm$^{-3}$; and
\item
double-detonation models from \citet{leung2020a} with model $M \approx
1~M_{\odot}$, $M_{\text{He}} \approx 0.1~M_{\odot}$, and spot- or
belt-like detonations.
\end{enumerate}

\subsection{Light Odd-$Z$ Elements}\label{light_odd_z}

Sodium, aluminum, and scandium are often referred to as light odd-$Z$
elements.  Like magnesium, sodium is mostly produced in core-collapse
supernovae via carbon burning.  Unlike magnesium, the surviving fraction
of sodium in supernovae ejecta depends on metallicity so it is treated
as a secondary product.  Sodium is also a product of hydrogen and helium
fusion in thermonuclear explosions, though in smaller quantities than
in core-collapse supernovae.  Similar to sodium, aluminum is synthesized
during carbon fusion in core-collapse supernovae in a secondary reaction
that is dependent on the amount of $^{22}$Ne burned (which in turn
depends on the carbon and oxygen content of the star).  While sodium
and aluminum are mostly produced in core-collapse supernovae, their
yields are dependent on metallicity and consequently their chemical
evolution is not as easily interpreted as the chemical evolution of
the $\alpha$-elements.  In contrast to sodium and aluminum, scandium
is formed via both oxygen burning in core-collapse supernovae and as a
product of $\alpha$-rich freeze-out in the shocked region just above the
rebounded core \citep{clayton2003}.  Both the exact nucleosynthetic origin
and chemical evolution of scandium is hard to precisely identify and
interpret because chemical evolution models largely underproduce scandium
\citep[e.g.,][]{clayton2003,zhao2016,prantzos2018,kobayashi2020a}.

Sodium and aluminum abundance inferences can be strongly affected
by departures from LTE, but scandium abundance inferences based on
\ion{Sc}{2} lines are not strongly affected \citep[e.g.,][]{zhao2016}.
We correct sodium abundances for departures from the assumptions
of LTE using the \citet{lind2011} grid through the INSPECT
project\footnote{\url{http://inspect-stars.com/}}.

\begin{figure*}
\plotone{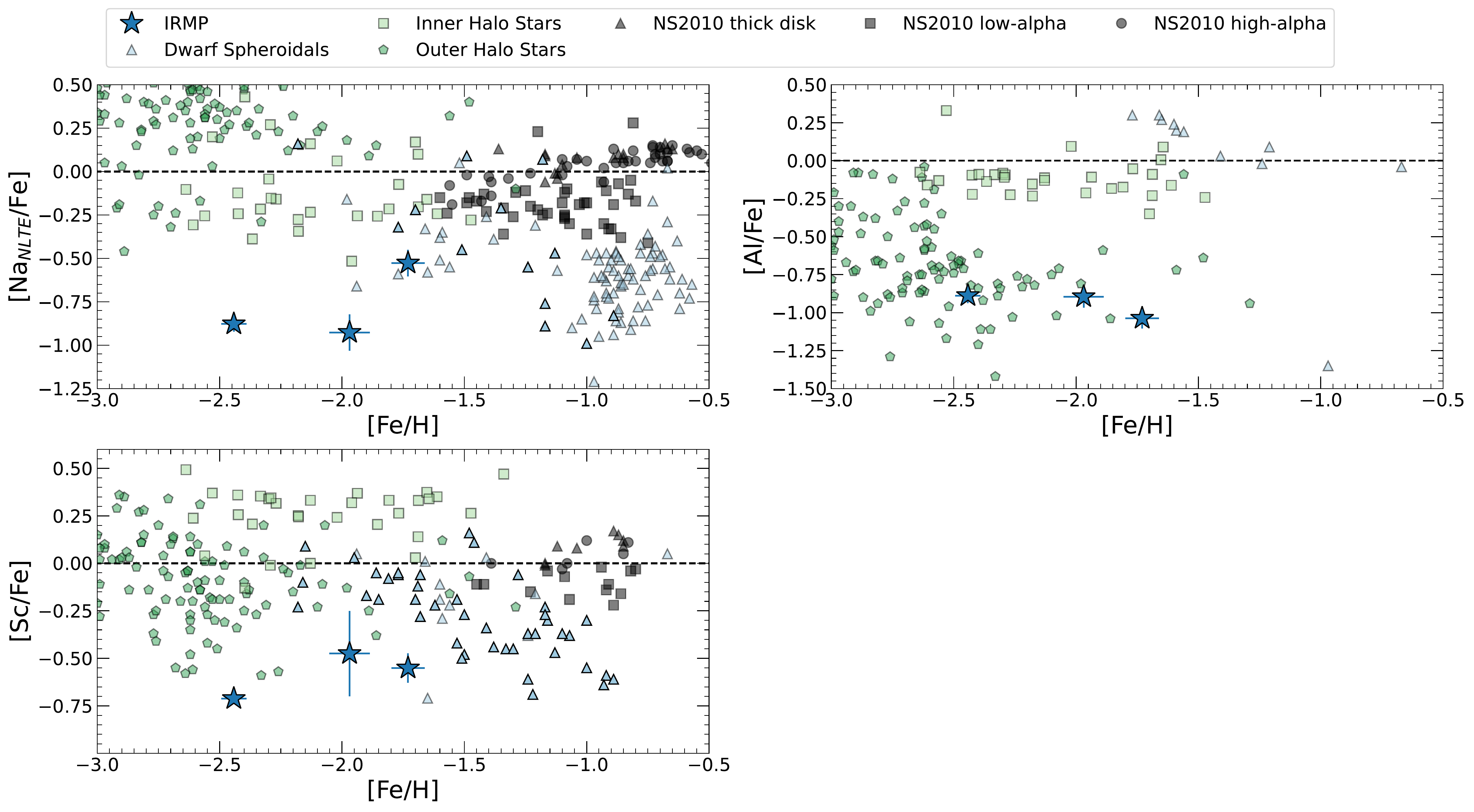}
\caption{Abundances of the light odd-$Z$ elements sodium, aluminum,
and scandium.  We plot as dark-blue stars the three iron-rich metal-poor
stars in our sample.  We plot as dark-blue stars the three iron-rich
metal-poor stars in our sample.   We plot as black-bordered light-green
squares inner halo metal-poor stars from \citet{reggiani2017} \&
Reggiani et al.\ (2023, in prep); as dark gray triangles, squares,
and circles thick disk, low-$\alpha$, and high-$\alpha$ stars from
\citet{nissen2010}; as black-boarded light-blue triangles stars from
the dwarf spheroidal galaxies Carina, Fornax, Sagittarius, and Sculptor
\citep{shetrone2003,geisler2005,monaco2005,letarte2010,hill2019,skuladottir2019};
and as dark-green pentagons outer halo metal-poor stars from
\citet{cayrel2004} \& \citet{jacobson2015}.  By construction, our three
IRMP stars have low $[\text{Na/Fe}]$ abundances.  Consistent with
the implications of subsolar $[\text{Ti/Fe}]$ abundances, subsolar
$[\text{Sc/Fe}]$ abundances generally disfavor double detonations in
thick helium envelopes.\label{light_odd_z_fig}}
\end{figure*}

We report in Table \ref{elem_abundances} the sodium, aluminum, and
scandium abundances we infer for the three IRMP stars in our sample and
plot them in Figure \ref{light_odd_z_fig} along with several comparison
samples.  While $[\text{Na/Fe}] < 0$ was required for these three IRMP
stars by construction, they also have subsolar $[\text{Al/Fe}]$ and
$[\text{Sc/Fe}]$ abundances.  The $[\text{Na/Fe}]$ and $[\text{Sc/Fe}]$
abundances of these three IRMP stars are at the lower envelope of the
$[\text{Na/Fe}]$ and $[\text{Sc/Fe}]$ distributions defined by our
comparison samples. We emphasize that some of our comparison
samples' sodium abundances we plot in Figure \ref{light_odd_z_fig}
have not been corrected for departures from the assumptions
LTE that according to \citet{lind2011} can be as large $-0.5$ dex
\citep[e.g.,][]{shetrone2003,cayrel2004,geisler2005,monaco2005,letarte2010,nissen2010}.
Because corrections for departures from the assumptions of LTE are
specific to individual lines as well as dependent on photospheric
parameters and equivalent width/LTE abundance, it would not be advisable
to apply them to previously published abundances.  Nevertheless, if we
did perform that exercise the sodium abundances we infer for the three
stars in our sample would still be at the lower envelope defined by our
comparison samples.  The observation of subsolar $[\text{Mg/Fe}]$,
$[\text{Ca/Fe}]$, $[\text{Na/Fe}]$, and $[\text{Sc/Fe}]$ abundances
in the same star is unusual in both the Milky Way and classical dSph
galaxies, suggesting that these three stars we define as IRMP stars
do indeed have usually larges amount of iron given their $\alpha$- and
light odd-$Z$ element abundances.  Consistent with the implications of
subsolar $[\text{Ti/Fe}]$ abundances, subsolar $[\text{Sc/Fe}]$ abundances
generally disfavor thermonuclear explosions produced by double detonations
in thick helium envelopes.

\subsection{Iron-peak Elements}\label{iron_peak}

Chromium, manganese, cobalt, nickel, and zinc are often referred to
as iron-peak elements.  Iron-peak elements can be formed directly in
or as a byproduct of explosive silicon burning, either incomplete
(chromium and manganese) or complete (cobalt, nickel, and zinc).
Their nucleosynthesis mainly takes place in thermonuclear explosions
\citep[e.g.,][]{clayton2003,grimmett2020}.  Cobalt and zinc yields are
correlated with the explosion energy of the nucleosynthetic event.
In this interpretation, the observed increase in the abundances of
$[\text{Co/Fe}]$ and $[\text{Zn/Fe}]$ with decreasing $[\text{Fe/H}]$
in metal-poor stars is evidence of an increased share of iron-peak
nucleosynthesis in core-collapse supernovae.  Contributions from
hypernovae events are possible at $[\text{Fe/H}] \lesssim -3.0$
\citep[e.g.,][]{cayrel2004,reggiani2017,kobayashi2020a}.  While chromium
abundance inferences based on \ion{Cr}{1} lines are strongly affected by
departures from the assumptions of LTE \citep[e.g.,][]{bergemann2010},
we also infer chromium abundances based on \ion{Cr}{2} lines that are
much less affected by departures from the assumptions of LTE.  Manganese,
cobalt, nickel, and zinc abundances inferences based on the lines we use
are not strongly affected by departures from the assumptions of LTE.
We correct iron abundances for departures from the assumptions of LTE
using data \citet{amarsi2016}.

\begin{figure*}
\plotone{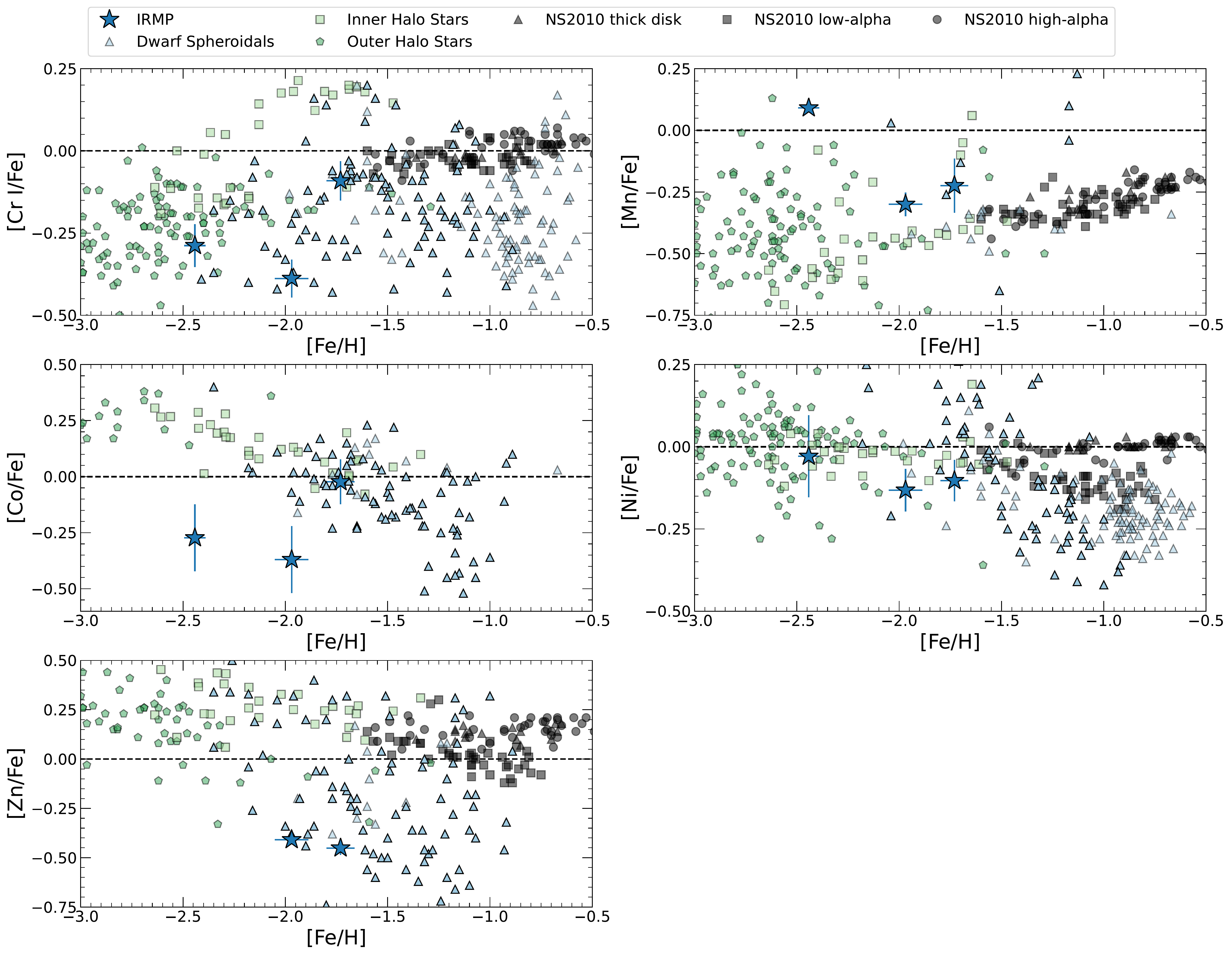}
\caption{Abundances of the iron-peak elements chromium, manganese,
cobalt, nickel, and zinc.  We plot as dark-blue stars the three iron-rich
metal-poor stars in our sample.  We plot as dark-blue stars the three
iron-rich metal-poor stars in our sample.   We plot as black-bordered
light-green squares inner halo metal-poor stars from \citet{reggiani2017}
\& Reggiani et al.\ (2023, in prep); as dark gray triangles, squares,
and circles thick disk, low-$\alpha$, and high-$\alpha$ stars from
\citet{nissen2010}; as black-boarded light-blue triangles stars from
the dwarf spheroidal galaxies Carina, Fornax, Sagittarius, and Sculptor
\citep{shetrone2003,geisler2005,monaco2005,letarte2010,hill2019,skuladottir2019};
and as dark-green pentagons outer halo metal-poor stars from
\citet{cayrel2004} \& \citet{jacobson2015}.  By construction, our three
IRMP stars have low $[\text{Co/Fe}]$.  The high $[\text{Mn/Fe}]$
abundances we infer support the idea that the iron-peak
elements in IRMP stars were mostly synthesized in thermonuclear
supernovae.\label{iron_peak_fig}}
\end{figure*}

We report in Table \ref{elem_abundances} the chromium, manganese,
cobalt, nickel, and zinc abundances we infer for the three IRMP
stars in our sample and plot them in Figure \ref{iron_peak_fig}
along with several comparison samples.  In accord with our comparison
samples, we plot in Figure \ref{iron_peak_fig} \ion{Cr}{1} abundances
despite their sensitivity to departures from the assumptions of LTE.
The $[\text{Co/Fe}]$ abundances of our three IRMP stars are all
subsolar by construction.  The $[\text{Cr/Fe}]$, $[\text{Ni/Fe}]$,
and $[\text{Zn/Fe}]$ abundances we infer for our sample are subsolar
as well, and our $[\text{Zn/Fe}]$ inferences for BD+80 245 and SMSS
J034249.53--284216.0 are at the lower envelope of the $[\text{Zn/Fe}]$
distribution defined by our comparison samples.  On the other hand,
the $[\text{Mn/Fe}]$ abundances of the three IRMP stars are at the upper
envelope of the $[\text{Mn/Fe}]$ distribution defined by our comparison
samples.

Almost all manganese is synthesized in thermonuclear explosions.
The high $[\text{Mn/Fe}]$ abundances we infer for our three IRMP stars
relative to our comparison samples support our assertion that IRMP stars
formed in environments where the contribution of thermonuclear events
to iron-peak nucleosynthesis was above average.  Indeed, the manganese
abundances of metal-poor stars has been used to study the detailed physics
of thermonuclear explosions \citep[e.g.,][]{reyes2020}.  Because cobalt
and zinc nucleosynthesis in core-collapse supernovae is correlated with
explosion energy, the low abundances $[\text{Co/Fe}]$ and $[\text{Zn/Fe}]$
in our three IRMP stars suggests powerful core-collapse supernovae did
not contribute much to the iron-peak abundances of our three IRMP stars.

\subsection{Neutron-capture Elements}\label{neutron_capture}

Elements with atomic numbers $Z \gtrsim 30$ like strontium, yttrium,
barium, and europium are often referred to as neutron-capture
elements because they are mostly synthesized through the capture
of neutrons by existing nuclei.  The neutron-capture timescale can
either be ``slow'' or ``rapid'' relative to $\beta$ decay timescales.
The relative contributions of these $s$- and $r$-processes to the
nucleosynthesis of each element are different and are functions
of metallicity.  Some elements like strontium, yttrium, and barium
are more commonly used as tracers of the $s$-process.  On the other
hand, europium is used as a tracer of $r$-process nucleosynthesis
\citep[e.g.,][]{cescutti2006,jacobson2013,ji2016a}.  Strontium, yttrium,
and europium abundances inferences based on the lines we use are not
strongly affected by departures from the assumptions of LTE.  We correct
barium abundances for departures from the assumptions of LTE using data
from \citet{amarsi2020}.

\begin{figure*}
\plotone{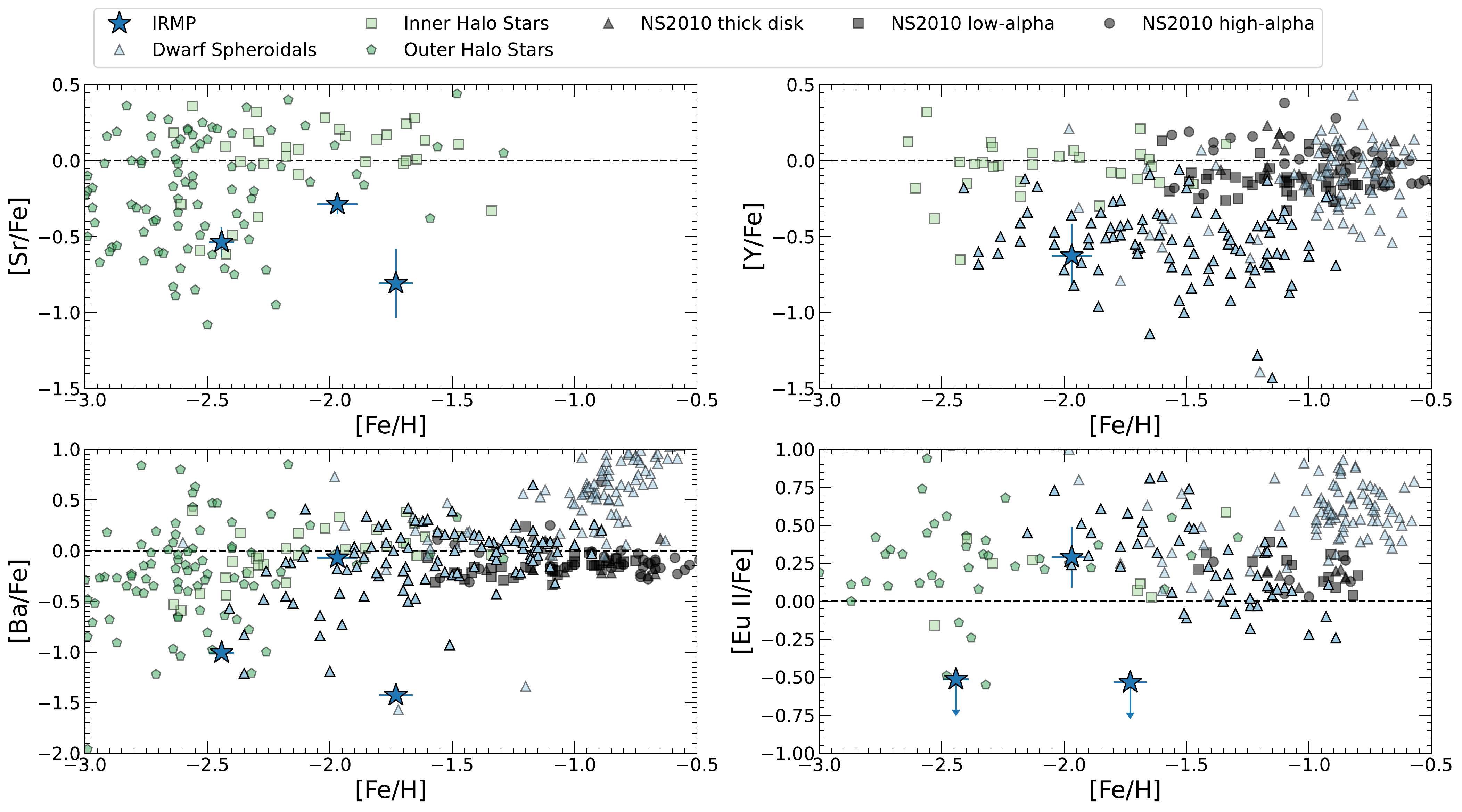}
\caption{Abundances of the neutron-capture elements strontium,
yttrium, barium, and europium.  We plot as black-bordered light-green
squares inner halo metal-poor stars from \citet{reggiani2017} \&
Reggiani et al.\ (2023, in prep); as dark gray triangles, squares,
and circles thick disk, low-$\alpha$, and high-$\alpha$ stars from
\citet{nissen2010}; as black-boarded light-blue triangles stars from
the dwarf spheroidal galaxies Carina, Fornax, Sagittarius, and Sculptor
\citep{shetrone2003,geisler2005,monaco2005,letarte2010,hill2019,skuladottir2019};
and as dark-green pentagons outer halo metal-poor stars from
\citet{cayrel2004} \& \citet{jacobson2015}.  As essentially
no neutron-capture elements are synthesized in thermonuclear
explosions, the low neutron-capture abundances relative to iron
we infer support the idea that the iron-peak elements in IRMP stars
were mostly synthesized in thermonuclear events without the additional
neutron-capture elements expected under more typical chemical evolution
scenarios.\label{neutron_capture_fig}}
\end{figure*}

We report in Table \ref{elem_abundances} the strontium, yttrium,
barium, and europium abundances we infer for the three IRMP stars in
our sample and plot them in Figure \ref{neutron_capture_fig} along
with several comparison samples.  In accord with our comparison
samples, we plot in Figure \ref{neutron_capture_fig} \ion{Sr}{2}
abundances that are insensitive to departures from the assumptions
of LTE \citep[e.g.,][]{hansen2013}.  At $[\text{Fe/H}] \gtrsim -2$,
the $[\text{Sr/Fe}]$ abundances we observe are very low.  Likewise,
the $[\text{Ba/Fe}]$ and $[\text{Eu/Fe}]$ abundances we inference
for BD+80 0245 and HE 0533--5340 are very low.  Since essentially no
neutron-capture elements are synthesized in thermonuclear explosions, the
low neutron-capture abundances relative to iron we see in our three IRMP
stars supports our assertion that IRMP stars formed in environments where
the contribution of thermonuclear events to iron-peak nucleosynthesis was
above average.  On the other hand, the star SMSS J034249.53--284216.0
has supersolar $[\text{Eu/Ba}]$, mildly enhanced $[\text{Eu/Fe}]$,
and highly enhanced $[\text{Eu/Mg}]$.  These three facts could possibly
result from a prolific $r$-process nucleosynthesis that occurred before
the thermonuclear event that produced most of its iron content.

\section{Discussion}\label{discussion}

We homogeneously analyzed the orbital properties, photospheric stellar
parameters, and elemental abundances of three giant stars BD+80 245,
HE 0533--5340, and SMSS J034249.53--284216.0 with $[\text{Fe/H}] < -1$
plus literature abundances $[\text{Na,Mg,Co/Fe}] < 0$ placing them in a
sparsely populated region of elemental abundance space consistent with
predictions for thermonuclear nucleosynthesis.  We found that these
three stars have have low-eccentricity orbits with apocenters inside
15 kpc typical of inner halo stars and therefore no orbital indication
that they formed in dwarf galaxies like the stars similarly studied
in \citet{reggiani2022a}.

Our state-of-the-art homogeneous abundance analyses confirmed that
these three stars have $[\text{Na,Mg,Al,Co,Zn,Sr,Ba/Fe}] < 0$ as
expected if they formed in environments dominated by thermonuclear
nucleosynthesis.  When compared to samples of inner halo, outer halo,
and classical dSph galaxy stars BD+80 245, HE 0533--5340, and SMSS
J034249.53--284216.0 have low $[\text{Mg/Fe}]$, $[\text{Ca/Fe}]$,
$[\text{Ti/Fe}]$, $[\text{Na/Fe}]$, $[\text{Sc/Fe}]$, $[\text{Co/Fe}]$,
$[\text{Zn/Fe}]$, $[\text{Sr/Fe}]$, and $[\text{Ba/Fe}]$ abundances.
They have high $[\text{Mn/Fe}]$ abundances.  BD+80 245 and HE 0533--5340
have very low upper limits $[\text{Eu/Fe}] \lesssim -0.5$, while SMSS
J034249.53--284216.0 has a more ordinary $[\text{Eu/Fe}] = 0.29$.  Because
most manganese and essentially no neutron-capture elements are synthesized
in thermonuclear explosions, the relatively high $[\text{Mn/Fe}]$
abundances combined with the relatively low $[\text{Sr/Fe}]$ and
$[\text{Ba/Fe}]$ abundances in these three stars support the idea that
they formed in environments dominated by thermonuclear nucleosynthesis.
Our results are consistent with previous conclusions based on similar
stars \citep[e.g.,][]{ivans2003,aoki2014,jeong2023}.  As will show though,
we are able to constrain specific thermonuclear explosion mechanisms.
We argue that these three stars are part of a class of metal-poor stars
we refer to as iron-rich metal-poor or IRMP stars.

The abundance properties shared between these three IRMP stars can be
used to explore the population of thermonuclear explosions.  The low
$[\text{Ca/Fe}]$ and $[\text{Ti/Fe}]$ abundances of these three IRMP
stars disfavor thermonuclear explosions produced by direct white
dwarf collisions or spot- or belt-like double detonations with either
especially thin or especially thick helium envelopes.  Similarly, the
low $[\text{Sc/Fe}]$ abundance of these three IRMP stars disfavors double
detonations in thick helium envelopes.

In an attempt to derive more specific constraints on the thermonuclear
explosions responsible for the abundance patterns we observe in these
three IRMP stars, we fit the grids of theoretical thermonuclear
supernovae yields from \citet{seitenzahl2013,seitenzahl2016},
\citet{fink2014}, \citet{ohlmann2014}, \citet{papish2016},
\citet{leung2018,leung2020a,leung2020b}, \citet{nomoto2018},
\citet{bravo2019}, \citet{boos2021}, \citet{gronow2021a,gronow2021b},
and \citet{neopane2022} to select the individual models that predict
abundances closest to those we observed in our sample.  Because both
core-collapse supernovae with delay times of a few Myr and thermonuclear
supernovae with delay times of a few tens of Myr contributed to the metal
contents of our three IRMP stars, we use a $\chi^{2}$-like statistic to
select the individual models that best predict our observed abundances.
Since we argue that a thermonuclear explosion produced the iron-peak
elements we observe in our three IRMP stars, we use the usual definition
of $\chi^{2}$ to assess the ability of the theoretical models listed
above to reproduce our observations.  On the other hand, one or more
core-collapse supernovae combined with a thermonuclear explosion to
produce the $\alpha$- and light odd $Z$-element abundance we observed.
As a result, we only penalize thermonuclear explosions that produce
$\alpha$- and light odd $Z$-element abundances beyond the level we
observed.  We exclude from our $\chi^{2}$-like statistic individual
elemental abundances for which a theoretical model predicts $\alpha$-
and light odd $Z$-element abundances below the level we observed.

To carry out the procedure described above, we first normalize the iron
yield of each model to the $[\text{Fe/H}]$ value we infer for each star.
We then identify the model that minimizes the modified $\chi^{2}$
statistic we define as
\begin{eqnarray}
\chi^{2} & = & \sum_{i} f([\text{X$_{i}$/Fe}]_{\text{o}},[\text{X$_{i}$/Fe}]_{\text{p}}) \frac{\left([\text{X$_{i}$/Fe}]_{\text{o}}-[\text{X$_{i}$/Fe}]_{\text{p}}\right)^{2}}{\sigma_{[\text{X$_{i}$/Fe}],\text{o}}} \nonumber \\
         &   & + \sum_{j} \frac{\left([\text{X$_{j}$/Fe}]_{\text{o}}-[\text{X$_{j}$/Fe}]_{\text{p}}\right)^{2}}{\sigma_{[\text{X$_{j}$/Fe}],\text{o}}},
\end{eqnarray}
where $[\text{X/Fe}]_{\text{o}}$ represents our observed abundances,
$[\text{X/Fe}]_{\text{p}}$ represents predicted abundances, and
$\sigma_{[\text{X/Fe}],\text{o}}$ represents the uncertainties in our
observed abundances.  The first summation is taken over the $\alpha$-
and light odd $Z$-elements sodium, magnesium, aluminum, silicon, calcium,
scandium, and titanium.  The second summation is taken over the iron-peak
elements chromium, manganese, cobalt, nickel, and zinc.  We define the
function $f([\text{X/Fe}]_{\text{o}},[\text{X/Fe}]_{\text{p}})$
\begin{equation}
f([\text{X/Fe}]_{\text{o}},[\text{X/Fe}]_{\text{p}}) \equiv
	\left\{ \begin{array}{ll}
	1 & \text{if}~[\text{X/Fe}]_{\text{o}} \leq [\text{X/Fe}]_{\text{p}}\\
        0 & \text{if}~[\text{X/Fe}]_{\text{o}} > [\text{X/Fe}]_{\text{p}}.
	\end{array} \right.
\end{equation}

\begin{figure*}
\plotone{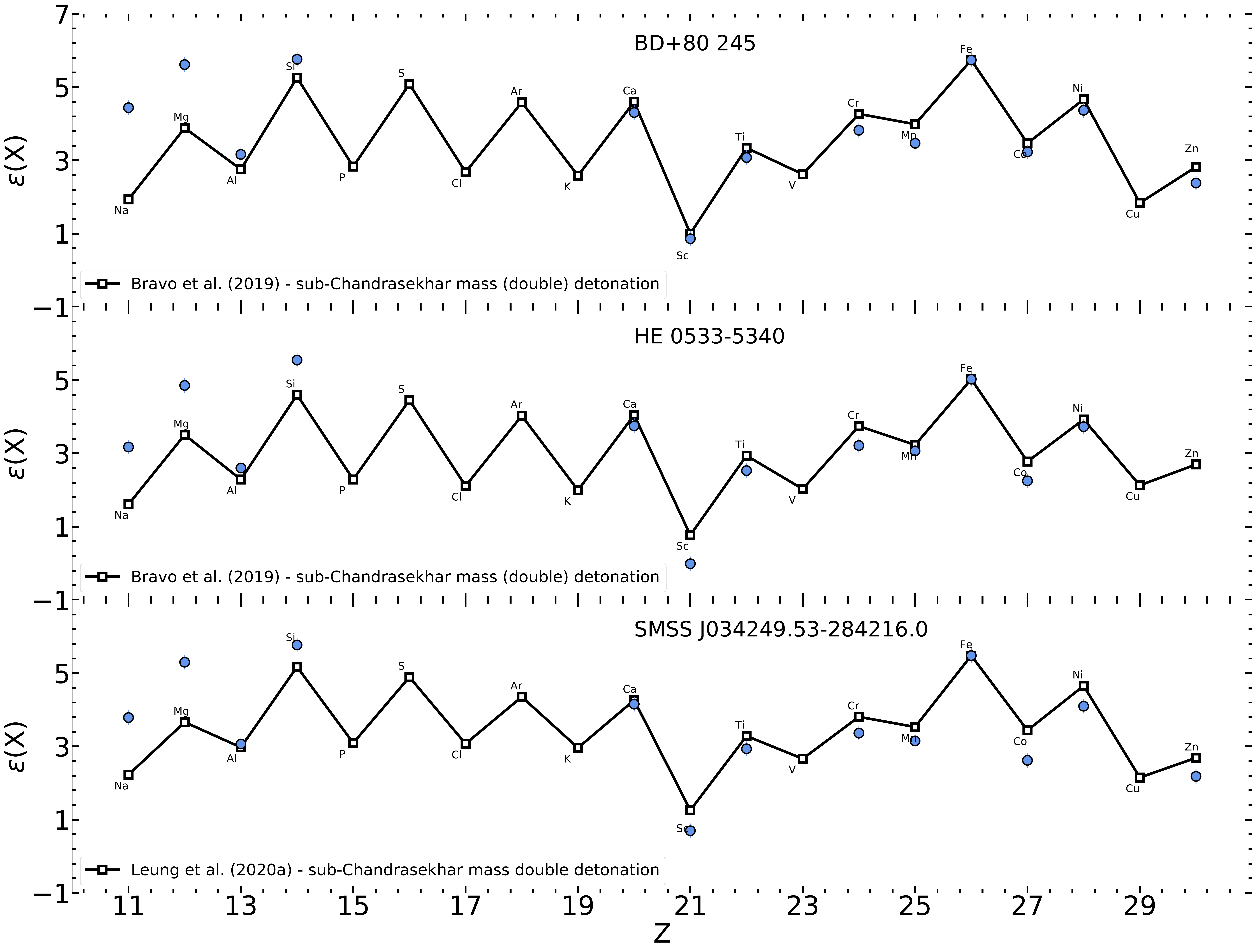}
\caption{Observed elemental abundances and predicted thermonuclear
event yields from models that best fit the observed data.  We plot as
blue circles our observed elemental abundances and as black-boarded
white squares connected by solid black lines the theoretical yields
predicted by the model that best reproduces our observations.  The best
model in each panel is depicted as black open squares, and the inferred
stellar abundances as blue circles.  The abundances of BD+80 245 and
HE 0533--5340 are best by \citet{bravo2019} models for the detonation
of sub-Chandrasekhar mass CO white dwarfs with $M_{\text{WD}} =
1.1~M_{\odot}$, a standard C/O ratio, and metallicities of $Z =
9 \times 10^{-3}$ and $Z = 2.25 \times 10^{-3}$ respectively.
The abundances of SMSS J034249.53--284216.0 are best fit by the
\cite{leung2020a} sub-Chandrasekhar mass double detonation model
110-050-2-S50.\label{progenitors_fig}}
\end{figure*}

We plot the abundances of BD+80 245, HE 0533--5340, and SMSS
J034249.53--284216.0 along with the models that best those abundances
in Figure \ref{progenitors_fig}.  We find that BD+80 245 and HE
0533-5340 are best fit by \citet{bravo2019} models for the detonation
of sub-Chandrasekhar mass CO white dwarfs with $M_{\text{WD}} =
1.1~M_{\odot}$, a standard C/O ratio, and metallicities of $Z = 9
\times 10^{-3}$ and $Z = 2.25 \times 10^{-3}$ respectively.  In the
one-dimensional \citet{bravo2019} models the detonation is arbitrarily
sparked at the center of the white dwarf.  This could be a spontaneous
pure detonation, a double detonation triggered by the detonation of a
thin surface helium shell, or a detonation resulting from the merger of
a double-degenerate system.  The abundances of SMSS J034249.53--284216
are best fit by the \cite{leung2020a} sub-Chandrasekhar mass double
detonation model 110-050-2-S50.

If our interpretation of IRMP stars is correct, then the occurrence of
IRMP stars should depend on galactic environment.  It is well established
that at constant $[\text{Fe/H}] \gtrsim -2.5$ surviving dSph
galaxies have lower $[\text{$\alpha$/Fe}]$ abundances than Milky Way
halo stars. The usual interpretation of this observation is
that the surviving dSph galaxies had extended star formation histories,
implying that thermonuclear nucleosynthesis was relatively more important
for the chemical evolution of dSph galaxies than for the Milky Way's halo
\citep[e.g.,][]{tolstoy2009,kirby2011a,kirby2011b,brown2012,brown2014,vargas2013,weisz2014}.
\citet{reggiani2021} showed that at $[\text{Fe/H}] \gtrsim -2$ the
same is true in the Magellanic Clouds.  On the other hand, the Milky
Way's globular clusters consistently have $[\text{$\alpha$/Fe}] \approx
0.4$, characteristically of core-collapse supernovae nucleosynthesis.
We assert that the occurrence of IRMP stars should be correlated with
the relative importance of thermonuclear nucleosynthesis to the chemical
evolution of a particular environment.  We predict that IRMP stars should
be more common in surviving dSph galaxies and the Magellanic Clouds than
in the Milky Way.  We further predict that IRMP stars should be rare in
globular clusters.

\section{Conclusions}\label{conclusion}

We collected published nucleosynthesis predictions for
thermonuclear explosions that could be responsible for the
astrophysical transients observationally classified as Type
Ia supernovae.  We find that thermonuclear explosion models
that yield $M_{\text{Fe}} \geq 0.1~M_{\odot}$ always produce
$[\text{C,N,O,F,Ne,Na,Mg,Al,Cl,K,Co,Cu,Zn/Fe}] < 0$, an abundance
space rarely populated by metal-poor stars with $[\text{Fe/H}] < -1$.
Focusing on the subset of these elements readily observable in the
photospheres of metal-poor giants stars $[\text{Na,Mg,Co/Fe}] < 0$, we
selected from the Stellar Abundances for Galactic Archaeology database
three metal-poor giants in this region of abundance space: BD+80 245,
HE 0533--5340, and SMSS J034249.53--284216.0.  We characterized the
Galactic orbits of these three stars and found no reason to believe that
they formed in now disrupted dwarf spheroidal galaxies.  We executed a
state-of-the-art, homogeneous abundance analysis for these three stars
and confirmed that they have $[\text{Na,Mg,Al,Co,Zn,Sr,Ba/Fe}] < 0$
as expected if they formed in environments dominated by thermonuclear
nucleosynthesis.  We argue that these three metal-poor stars BD+80 245,
HE 0533--5340, and SMSS J034249.53--284216.0 should be considered the
first examples of a new class of metal-poor stars that we refer to
iron-rich metal-poor (IRMP) stars that formed in environments dominated
by thermonuclear nucleosynthesis.  The elemental abundances of these
three stars disfavor thermonuclear explosions produced by direct white
dwarf collisions or double detonations in either very thin or very thick
helium shells.  The elemental abundances of BD+80 245, HE 0533--5340, and
SMSS J034249.53--284216.0, are best explained by the (double) detonation
of sub-Chandrasekhar mass CO white dwarfs.  If our interpretation of
IRMP stars is correct, then then they should be very rare in globular
clusters and more common in the Magellanic Clouds and dwarf spheroidal
galaxies than in the Milky Way’s halo.

\section*{Acknowledgments}

We thank Ian Roederer, Ian Thompson, and Heather Jacobson for providing us
their spectra of BD+80 245, HE 0533--5340, and SMSS J034249.53--284216.0.
We thank Abigail Polin and Yossef Zenati for numerous helpful discussions
about the physics of thermonuclear explosions and the observational
properties of Type Ia supernovae.  Henrique Reggiani acknowledges
support from a Carnegie Fellowship.  Parts of this research were
supported by the Australian Research Council Centre of Excellence
for All Sky Astrophysics in 3 Dimensions (ASTRO 3D), through project
number CE170100013. The national facility capability for SkyMapper
has been funded through ARC LIEF grant LE130100104 from the Australian
Research Council, awarded to the University of Sydney, the Australian
National University, Swinburne University of Technology, the University
of Queensland, the University of Western Australia, the University of
Melbourne, Curtin University of Technology, Monash University and the
Australian Astronomical Observatory.  SkyMapper is owned and operated
by The Australian National University's Research School of Astronomy
and Astrophysics.  The survey data were processed and provided by
the SkyMapper Team at ANU.  The SkyMapper node of the All-Sky Virtual
Observatory (ASVO) is hosted at the National Computational Infrastructure
(NCI). Development and support of the SkyMapper node of the ASVO has been
funded in part by Astronomy Australia Limited (AAL) and the Australian
Government through the Commonwealth's Education Investment Fund (EIF)
and National Collaborative Research Infrastructure Strategy (NCRIS),
particularly the National eResearch Collaboration Tools and Resources
(NeCTAR) and the Australian National Data Service Projects (ANDS).
This work has made use of data from the European Space Agency (ESA)
mission {\it Gaia} (\url{https://www.cosmos.esa.int/gaia}), processed
by the {\it Gaia} Data Processing and Analysis Consortium (DPAC,
\url{https://www.cosmos.esa.int/web/gaia/dpac/consortium}).  Funding for
the DPAC has been provided by national institutions, in particular the
institutions participating in the {\it Gaia} Multilateral Agreement.
This publication makes use of data products from the Two Micron All
Sky Survey, which is a joint project of the University of Massachusetts
and the Infrared Processing and Analysis Center/California Institute of
Technology, funded by the National Aeronautics and Space Administration
and the National Science Foundation.  This publication makes use of
data products from the Wide-field Infrared Survey Explorer, which is
a joint project of the University of California, Los Angeles, and the
Jet Propulsion Laboratory/California Institute of Technology, funded by
the National Aeronautics and Space Administration.  This research has
made use of the NASA/IPAC Infrared Science Archive, which is funded by
the National Aeronautics and Space Administration and operated by the
California Institute of Technology.  This research has made use of the
SIMBAD database, operated at CDS, Strasbourg, France \citep{wenger2000}.
This research has made use of the VizieR catalogue access tool,
CDS, Strasbourg, France (DOI: 10.26093/cds/vizier).  The original
description of the VizieR service was published in 2000, A\&AS 143,
23 \citep{ochsenbein2000}.  This research has made use of the Keck
Observatory Archive (KOA), which is operated by the W. M. Keck Observatory
and the NASA Exoplanet Science Institute (NExScI), under contract with
the National Aeronautics and Space Administration.  This research has
made use of NASA's Astrophysics Data System Bibliographic Services.


\vspace{5mm}
\facilities{ADS, CDS, CTIO:2MASS, FLWO:2MASS, Gaia, GALEX, KOA, IRSA,
Magellan:Clay, NEOWISE, Skymapper, Smith, WISE}

\software{\texttt{astropy} \citep{astropy2013,astropy2018},
          \texttt{colte} \citep{casagrande2021},
          \texttt{isochrones} \citep{morton2015},
          \texttt{MOOG} \citep{sneden1973},
          \texttt{MultiNest} \citep{feroz2008,feroz2009,feroz2019},
          \texttt{numpy} \citep{harris2020},
          \texttt{pandas} \citep{mckinney2010,reback2020},
          \texttt{PyMultinest} \citep{buchner2014},
          \texttt{q$^{2}$} \citep{ramirez2014},
          \texttt{R} \citep{r22},
          \texttt{scipy} \citep{virtanen2020}
          }

\bibliography{ms}{}
\bibliographystyle{aasjournal}

\end{document}